\documentclass[12pt,preprint]{aastex}

\def\gsim{\;\rlap{\lower 2.5pt\hbox{$\sim$}}\raise 1.5pt\hbox{$>$}\;}
\def\lsim{\;\rlap{\lower 2.5pt\hbox{$\sim$}}\raise 1.5pt\hbox{$<$}\;}
\def\del{{\partial}}
\newcommand{\HDblah}{HD~$\!$209458~$\!$b}
\newcommand{\LR}{$L_{\cal R}$}

\shorttitle{CIRCULATION OF CLOSE-IN EXTRASOLAR GIANT PLANETS I}
\shortauthors{CHO, MENOU, HANSEN \& SEAGER} 

\begin{document}

\title{Atmospheric Circulation of Close-In Extrasolar Giant Planets:\\
  I.\, Global, Barotropic, Adiabatic Simulations}

\author{James Y-K.\ Cho,}

\affil{Department of Terrestrial Magnetism, Carnegie Institution of
  Washington,\\ 5241 Broad Branch Road, N.W., Washington, D.C. 20015,
  USA \\
  and \\
  Astronomy Unit, School of Mathematical Sciences, 
  Queen Mary, University of London, \\
  Mile End Road, London E1 4NS, UK}

\author{Kristen Menou,}
 
\affil{Department of Astronomy, Columbia University,\\ 550 W. 120th
  Street, New York, NY 10027, USA}

\author{Bradley M.\ S.\ Hansen,}

\affil{Department of Physics and Astronomy and Institute for
  Geophysics and Planetary Physics, University of California,
  475 Portola Plaza, Box 951547, Los Angeles, CA 90095, USA}

\and

\author{Sara Seager}
 
\affil{Department of Terrestrial Magnetism, Carnegie Institution of
  Washington,\\ 5241 Broad Branch Road, N.W., Washington, D.C. 20015,
  USA}

\begin{abstract}
  We present results from a set of over 300 pseudospectral simulations
  of atmospheric circulation on extrasolar giant planets with circular
  orbits.  The simulations are of high enough resolution (up to 341
  total and sectoral modes) to resolve small-scale eddies and waves,
  required for reasonable physical accuracy.  In this work, we focus
  on the global circulation pattern that emerges in a shallow,
  ``equivalent-barotropic'', turbulent atmosphere on both tidally
  synchronized and unsynchronized planets.  A full exploration of the
  large physical and numerical parameter-space is performed to
  identify robust features of the circulation.  For some
  validation, the model is first applied to Solar System giant
  planets.  For extrasolar giant planets with physical parameters
  similar to \HDblah---a presumably synchronized extrasolar giant
  planet, representative in many dynamical respects---the circulation
  is characterized by the following features: 1) a coherent polar
  vortex that revolves around the pole in each hemisphere; 2) a low
  number---typically two or three---of slowly-varying, broad zonal
  (east-west) jets that form when the maximum jet speed is comparable
  to, or somewhat stronger than, those observed on the planets in the
  Solar System; and, 3) motion-associated temperature field, whose
  detectability and variability depend on the strength of the net
  heating rate and the global root mean square wind speed in the
  atmosphere.  In many ways, the global circulation is Earth-like,
  rather than Jupiter-like.  However, if extrasolar giant planets
  rotate faster and are not close-in (therefore not synchronized),
  their circulations become more Jupiter-like, {for Jupiter-like
  rotation rates}.
\end{abstract}


\section{Introduction}

A large number of extrasolar planets orbiting nearby sun-like stars is
now known.\footnote{See, e.g., {\tt{\url
      http://exoplanets.org/almanacframe.html}} and {\tt {\url
      http://www.obspm.fr/encycl/encycl.html}}.}  So far, all of the
planets are giant planets, which orbit $\!\lsim\! 5$~AU from their
host stars.  These extrasolar giant planets possess very interesting
and unexpected orbital properties.  Some of them orbit extremely close
to their host stars (semi-major axis $a \lsim$0.1~AU) and have very
nearly circular orbits (eccentricity $e \lsim 0.05$).  These are the
close-in extrasolar giant planets.  If they possess tidal $Q$
parameter values similar to that of Jupiter ($\sim$10$^6$), they are
expected to be in 1:1~spin-orbit resonance, or {\it synchronous},
states \citep{Goldreich66,Rasio96,Lubow97,Ogilvie04}.  The rest of the
extrasolar giant planets orbit further away from their host stars,
generally with substantial eccentricities.  Among them, some have
nearly circular orbits however, like the Solar System giant planets.
In this paper, we focus our study on the atmospheric circulation of
all giant planets with circular orbits---from the close-in extrasolar
giant planets to the more distant ``unsynchronized'' extrasolar giant
planets, as well as the Solar System giant planets.
  
The existence of a dynamic, stably-stratified (radiative) outer layer
is a common feature on planets with atmospheres.  Extrasolar giant
planets are also expected to possess such layers---independent of the
proximity to their host stars.  The dynamic layer is typically
$\lsim\! 10$ pressure scale heights thin, much smaller than the radius
of the planet.  It extends from just below the ``weather layer'' (at
the bottom) up to the turbopause (at the top)---i.e., from near the
top of the convection zone up to the tenuous region where molecular
diffusion begins to dominate over advection.  Following the first
detections of secondary eclipses for three transiting extrasolar giant
planets by the {\it Spitzer Space Telescope}
\citep{Charbonneau05,Deming05,Deming06}, the radiative layers of
extrasolar giant planets are now the focus of intense observational
and theoretical efforts
\citep{Seager05,Barman05,Burrows05,Burrows06,Fortney05,Iro05,Williams06}.
Since vigorous motions and meteorology in these layers will strongly
influence observable properties, a good understanding of atmospheric
dynamics on extrasolar giant planets is needed.  The understanding is
especially crucial for the synchronized planets with their unique
heating configuration.  It is uncertain what temperature distribution
should result from the combined effects of stationary dayside heating
and advection on the atmospheres of these planets
\citep[e.g.][]{Showman02,Cho03}.  This is partly because they are
dynamically very different from the Solar System giant planets
\citep{Cho03,Menou03}.  Close-in extrasolar giant planets are also
independently important for theory since their expected synchronous
states provide an idealized forcing situation, important for improving
understanding of planetary (as well as possibly stellar) atmospheres
in general.

Recently, several global simulations of close-in extrasolar giant
planet atmospheric circulation have been performed, focusing on
different aspects of the overall complex problem
\citep{Showman02,Cho03,Cooper05}.  Global calculations are necessary
to study the full effect of rotation on large-scale\footnote{By
  ``large-scale'', it is meant horizontal scale $L\gsim R_p/10$, where
  $R_p$ is the radius of the planet.} flows---the general
circulation---on planets.  \citet{Cho03} have carried out a
high-resolution, one-layer turbulence simulation of the close-in
extrasolar giant planet \HDblah\ using the pseudospectral method.
They stress that the gross feature of the atmospheric flow on this
``Hot Jupiter'' is markedly different from that on {\it the} Jupiter:
the horizontal circulation pattern is actually more like that of the
Earth or Venus, near and above the cloudtops.  As elucidated in the
companion paper by \citet{Menou03}, this is expected on dynamical
grounds.  The marked difference is due to the close-in planet's slower
rotation rate ($\Omega^{\rm HD}\approx 2.1\!\times\!
10^{-5}$~rad~s$^{-1}$) and hotter equilibrium temperature ($T_e^{\rm
  HD}\!\sim\! 1500$~K), compared to those of Jupiter ($\Omega^{\rm
  Jup} = 1.4 \times$10$^{-4}$~rad~s$^{-1}$ and $T_e^{\rm Jup} =
130$~K).  \citet{Showman02} and \citet{Cooper05} have also performed
global simulations.  Using the finite difference method in
multiple-layer models, they focus on the laminar state and the
vertical structure of the circulation.  Both of these multiple-layer
studies solve a more general set of equations than that of
\cite{Cho03}, allowing vertical variations over many pressure scale
heights to be represented.  The multiple-layer studies also differ in
their model setup: in most cases, an atmosphere {\it at rest} is
heated, leaving out the {important} influences of
{pre-existing} jet streams and {evolving} turbulent
eddies.

The structure and robustness of {well-resolved horizontal} flow
forming under applied heating is the primary subject of this paper.
Specifically, global circulation patterns possible in a
stably-stratified, turbulent atmosphere on extrasolar giant planets
with circular orbits are studied.  This paper presents several
much-needed developments in extrasolar planet characterization study:
1)~a broad parameter-space exploration of a common type of extrasolar
giant planet---i.e., extrasolar giant planets with circular orbits,
which include the transiting close-in extrasolar giant planets; 2) a
validation of the circulation model against a previously successful
giant planet circulation model; and, 3) a detailed description of the
employed model so that equitable comparisons can be made with
alternative flow models.  As in \citet{Cho03}, we here use the
``equivalent-barotropic formulation'' \citep{Salby89} of the {\it
  primitive equations} of meteorology (cf., \citet{Holton92} and
\citet{Salby96}), hereafter referred to as the {\it
  equivalent-barotropic equations}.  The equivalent-barotropic
equations allow us to practically perform the large number of
high-resolution simulations necessary for the exploration while
bypassing the issue of the current lack of information on the vertical
structure (e.g., distribution of radiatively-active species).  The
equivalent-barotropic equations also allow a physically clear
interpretation of heating and cooling associated with dynamics in a
stratified layer.  In this paper, we focus on the adiabatic
dynamics---appropriate for when the dynamical timescales are shorter
than the radiative timescales or when diabatic heating and cooling
nearly cancel in the net.  The focus sets the stage for future
diabatic (non-adiabatic) calculations.

The overall plan of the paper is as follows.  In \S2, we briefly
describe the current state of understanding of giant planet
atmospheric circulation, laying stress on the aspects that are
important for extrasolar giant planets.  In \S3, we present the
primitive equations, the equations that govern the large-scale
dynamics of atmospheres, and discuss the origins and properties of the
reduced version of the primitive equations, the equivalent-barotropic
equations.  Readers wishing more details on the derivation of the
equivalent-barotropic equations are referred to \citet{Salby89}.
Readers familiar with the equations of atmospheric motion may skip the
early part of this section.  Later in the same section, we describe
our representation of the differential thermal forcing, important for
modeling close-in extrasolar giant planet atmospheres.  In \S4, we
present the simulations of Solar System giant planets using the
equivalent-barotropic equations.  The simulations provide model
validation as well as an initial assessment of circulation on
unsynchronized extrasolar giant planets, which are essentially modeled
in the present work as Solar System giant planets with greater
equilibrium temperature $T_e$ and/or slower rotation rate~$\Omega$.
In \S5, we summarize the detailed parameter-space exploration of the
circulation on \HDblah, a representative close-in extrasolar giant
planet in many dynamical respects.  This brackets circulations
possible on close-in extrasolar giant planets in the context of
one-layer modeling.  In \S6, we conclude our presentation of the
global adiabatic calculations.

\section{Atmospheric Circulation on Giant Planets}

The dynamically-active layer of a giant planet atmosphere is forced by
agents both external and internal to the layer, which render the
large-scale flow in the layer very complex.  The forcing comes from
{\it i}) the host star irradiation above, {\it ii}) the convection
below, and {\it iii}) the momentum and heat fluxes within.  The
impinging stellar radiation is mostly shortwave (i.e., ultraviolet and
visible).  Some of the incoming radiation is scattered by the
atmospheric gases, which is composed mostly of inert H$_2$ and He by
volume.  Some of it is reflected back by clouds, if they are present.
And, some of it is absorbed by the atmosphere, particularly by active
minor constituents such as H$_2$O vapor and clouds.  The atmospheric
gases and clouds also emit and absorb longwave (i.e., infrared)
radiation, leading to either further heat transfer between different
parts of the atmosphere or heat loss to space.  Convection provides
direct stirring as well as latent heat from condensation.  Roughly
speaking, the static atmospheric structure is in radiative equilibrium
above the weather layer (in the middle and upper part of the active
layer) and in radiative-convective equilibrium below the weather layer
(near the bottom of the active layer). In this way, radiative transfer
processes are important in establishing the vertical temperature
structure throughout the active layer.  However, dynamical processes
(such as transport and dissipation of eddies\footnote{ ``Eddies'' are
  the residual of the dynamical field after the subtraction of the
  mean.  In geophysical fluid dynamics, ``eddies'' are often used
  interchangeably with ``vortices'' and even ``waves''.} and waves)
significantly modify the basic temperature structure that would be
established by the radiative processes alone, in the absence of
motion.

On Solar System giant planets, despite the markedly different orbital,
thermal, and chemical properties, the atmospheres all exhibit strong
zonal (east-west) banding and associated zonal jets
(longitudinally-averaged eastward winds), with long-lasting vortices
embedded in them.  The bands, jets, and vortices can all be understood
as a dynamical equilibrium state of a stably-stratified shallow layer
of turbulent fluid on a rotating sphere \citep{Cho96a}.  In a
stratified-rotating fluid, the motion is predominantly layer-wise and
two-dimensional (2-D).  As a consequence, the flow exhibits a
simultaneous inverse cascade of energy up to the large scales and
forward cascade of enstrophy\footnote{Enstrophy is the ``vortical
  energy'' of the fluid, defined by $\onehalf\zeta^2$, where
  $\vec\zeta = \zeta{\bf k} = \nabla\!\times\!{\bf v}$ is the
  vorticity (parallel to the vertical direction ${\bf k}$) and ${\bf
    v}\in\Re^2$ is the horizontal velocity.}  down to the small scales
\citep{Kraichnan67,Leith68,Batchelor69}.  This is in marked contrast
to the usual forward cascade of energy (only) in 3-D turbulence
\citep{Kolmogorov41a,Kolmogorov41b,Kolmogorov41c}.  In 2-D turbulence,
the inverse cascade roughly corresponds to growth of eddies (vortices)
by successive mergers \citep{McWilliams84}. In an isotropic condition,
the eddies ultimately grow to the size of the domain.  On the sphere,
however, a fundamental anisotropy due to the geometry is present, if
the sphere is rotating: Coriolis acceleration provides a restoring
force on the growing eddies in the meridional (north-south) direction
while none in the zonal direction.  In this anisotropic condition,
eddies are free to grow to the largest available scale in the zonal
direction but restricted in the meridional direction---thus forming
the bands and jets \citep{Rhines75}.  In this heuristic explanation of
the bands, the characteristic meridional width of the bands is loosely
defined by the Rhines scale, $L_\beta\equiv\pi(2U/\beta)^{1/2}$, where
$U$ is the root mean square wind speed of the jets and $\beta\equiv
2\Omega\cos\phi / R_p$ with $\phi$ the latitude and $R_p$ the radius
of the planet.  According to this mechanism, the number of jets and
bands is $\approx\!\pi R_p/L_\beta$ and is an increasing function of
the planetary rotation rate $\Omega$.

\citet{Cho96b} have explicitly shown in numerical simulations that the
banded structures on all Solar System giant planets can be reproduced
very well through this ``$\beta$-mechanism''.  In their study, they
use the {\it shallow-water equations}, which formally applies to a
thin layer of water with a free surface \citep{Pedlosky87}.  The
shallow-water simulations capture reasonably well the number, size,
and amplitude of the zonal jets---the gross features of the
atmosphere---on all four giant planets of the Solar System.  In
addition, the model captures the anticyclonic\footnote{Cyclones
  (anticyclones) are vortices defined by the positive (negative) sign
  of $\vec\zeta\cdot{\bf \Omega}$, where ${\bf\Omega}$ is the
  planetary rotation vector (and defines the north pole).} dominance
on Jupiter \citep{MacLow86}.  It is important to note that their
simulations use only the {\it observed} physical parameters values of
$\Omega$, $g$, $R_p$, $\bar{U}$, and $H_c$ (respectively, planetary
rotation rate, surface gravity, planetary radius, global root mean
square wind speed, and thickness based on the scale height at the
cloudtop level).
The qualitative match obtained between the simulations and the
observations gives confidence in the basic approach of using a
barotropic, adiabatic, shallow-layer turbulence model to study giant
planet atmospheric circulations.  Similar results using the
equivalent-barotropic equations is discussed in
\S\ref{sec:ssebf}. Alternative models of bands and jets on a giant
planet also exist, however.  One model is based on the surface
expression of putative deep, convective, Taylor-Proudman columns
aligned with the planetary rotation vector ${\bf \Omega}$ (cf.,
\citet{Sun93}, and references therein).
Recently, \citet{Heimpel05} have also obtained a good match to
Jupiter's surface zonal wind structure using a 3-D global convection
model.  Their calculation is performed in a thin spherical zone, which
in principle is similar to the shallow layer turbulence approach
(rather than the deep convection laminar approach).  

Given the relative success of the shallow-water model\footnote{We must
keep in mind, however, the gross simplifications inherent in the
shallow-water model (see \S3).  The model does not---and in many cases
cannot---address a number of important aspects of the circulation.},
it is natural to try and extend the {barotropic} approach to
extrasolar giant planets.  However, a number of issues critical for
the study of extrasolar giant planet atmospheric dynamics is currently
either poorly understood or await observational constraints.  This
entails that current extrasolar planet atmosphere models, including
radiative transfer models, necessarily contain some basic and crucial
unknowns.  The vertical distribution of active species has already
been mentioned; $\Omega$ is another.  The latter {is} crucial
in characterizing the planet's visual, spectral, and hydrodynamic
properties \citep{Cho96b,Showman02,Cho03,Menou03}.  Since orbital
periods can be accurately measured, the rotation rate of close-in
extrasolar giant planets can at least be inferred by assuming
synchronization.  However, even in this case, there is still the
question of how to initialize the model calculation.  A background
zonal flow, for example, will nontrivially affect the evolution of any
developing flow by nonlinearly interacting with it.  Without adequate
initial and boundary conditions, it cannot be stressed enough that
current model calculations should more properly be considered as
revealing important mechanisms and plausible flow states, rather than
predicting reality.

With the above caveats in mind, we forge ahead by identifying giant
planet properties that should be general.
The flow in the radiative layer is generally not driven by the stellar
heating directly but by the deviation from the equilibrium
temperature.  Moreover, the deviation responsible for the transport is
{\it caused} by the atmospheric dynamics---specifically, by eddies and
waves.  Hence, it is more proper to think of the dynamics as
controlling the temperature distribution, rather than the other way
around.  This is why it is crucial to represent the flow dynamics
properly, both initially and throughout the flow evolution.
Well-known examples of eddy/wave-induced temperature and flow
modifications exist.  Rossby (``planetary'') and gravity (buoyancy)
waves cause major warmings over the cold pole, forcing the region to
be above the radiative equilibrium temperature and undergo cooling.
Waves can also induce quasi-periodic flow, and its associated
temperature reversals, in the equatorial region.  Cold regions at the
subsolar point just above the troposphere and at the top of the
radiative layer are eddy/wave-driven as well, as the rising air in
those regions cool adiabatically.  These are fairly well understood
phenomena in the Earth's atmosphere, where considerable observational
data is available (cf., \citet{Andrews87}).  Remarkably, similar
heating and cooling mechanisms appear likely to operate on extrasolar
giant planets, as we shall show later (\S4 and \S5).  On
rapidly-rotating and unsynchronized extrasolar giant planets,
meridional transfer of heat is expected; and, on slowly-rotating or
synchronized extrasolar giant planets, zonal transfer of heat is also
expected, in addition to the meridional transfer.

Under certain circumstances, Newtonian cooling approximation may be
fruitfully used to represent the net heating (cf., \citet{Andrews87}
and \citet{Salby96}).  The approximation is a simple representation of
radiative effects on large-scale dynamics---a relaxation of the
temperature field to a specified ``equilibrium'' distribution, which
in fact depends in a complicated way on the atmospheric constituents
and their spatial distributions.  It is not known {\it a priori} in
general.  The approximation is crude in that it ignores vertical
motion ($dp/dt$), assumes maximum temperature perturbation
$T^{\prime}$ small compared to the equilibrium temperature (i.e.,
$T^{\prime}\ll\ $1500~K for close-in extrasolar giant planets),
approximates the vertical gradient of the transmission function to be
$\sim$1/$H_p$, and requires the environment to be in local
thermodynamic equilibrium (cf., \citet{Andrews87}).  It is widely used
in theoretical studies of the Earth's atmospheric dynamics, but
caution should be exercised when {liberally} extending its use
to extrasolar giant planets.

In this paper, we focus on the dynamical implications of a specified
temperature distribution containing a large, steady day-night
component.  In the context of the Newtonian cooling approximation, the
relevance of this steady regime can be understood in terms of the
relative magnitude of the advective timescale, $\tau_a\!\sim\!L/U$,
and radiative timescale, $\tau_r\sim \rho c_p(d{\cal
  T}/dz^*)^{-1}/(4\sigma T^3)$.  Here, $L$, $U$, $p$, $c_p$, $\sigma$,
$T$, ${\cal T}$, and $z^*$ are the characteristic length,
characteristic speed, basic density, specific heat at constant
pressure, Stefan-Boltzmann constant, equilibrium temperature,
transmission function, and log-pressure height ($=\!  H_p\ln\,(p_{\rm
  ref}/p)$, with $H_p$ the pressure scale height and $P_{\rm ref}$ a
reference pressure), respectively.  When $\tau_r \gg \tau_a$,
advection of heat is much faster than radiative heat exchanges and the
dynamics is essentially adiabatic.  Based on preliminary diabatic
calculations we have performed with explicit Newtonian cooling, the
cases when $\tau_r > \tau_a$ in our simulations are not qualitatively
very different from the $\tau_r\!\gg\!\tau_a$ case---unless
$\tau_r\!\lsim\!\tau_a$.  Hence, adiabatic simulations are relevant
over a wide range of $\tau_r$.  In the opposite limit, when
$\tau_r\!\ll\!  \tau_a$, the flow dynamics equations cease to be
{physically} connected to the fast radiative forcing.
As we argue in \S3.4, in this case it is reasonable to model this
quasi-instantaneous radiative forcing limit by imposing a fixed
temperature structure.  {A detailed study, which focuses on the
  relaxation of this model constraint, will be presented elsewhere.}

\section{Circulation Model}

\subsection{The Primitive Equations}

The motion of an atmospheric fluid element is governed by the
primitive equations.  The equations are so called because they
constitute the starting point for studying large-scale dynamics in the
atmosphere.  The presumption in their use is that the small scales,
which are not resolvable either observationally or numerically, are
{\it parameterizable} within the framework of large-scale dynamics.
In atmospheric studies, $p$ is commonly used as the vertical
coordinate, rather than the geometric height, $z = z({\bf x},p,t)$
with ${\bf x}\!\in\!\Re^2$; $z$ becomes a dependent variable in this
case.  In the standard $p$-coordinates, the primitive equations are:
\begin{mathletters}\label{primitive equations}
\begin{eqnarray}
\frac{D{\bf v }}{D t} + 
\frac{u\tan\phi}{R_p}\, {\bf k}\!\times\!{\bf v}\! &\! =\! &\!
-\nabla\! _p\ \Phi - f{\bf k}\!\times\!{\bf v} + {\cal F} - {\cal D}\ ,\\
\frac{\del\Phi}{\del p}\! &\!  =\! &\! -\alpha\ ,\\
\frac{\del\omega}{\del p} &\! =\! &\! -\nabla\! _p\cdot{\bf v}\ ,\\ 
\frac{D\theta}{D t} &\! =\! &\! \frac{\theta}{c_p T}\ \dot{q}_{\rm
  net}\ ,
\end{eqnarray}
\mbox{where}
\begin{equation}
\frac{D}{D t} \ = \ \frac{\del}{\del t} + 
{\bf v}\!\cdot\!\nabla\! _p + \omega\!\frac{\del}{\del p}\ .
\end{equation}
\end{mathletters}\noindent
In Eq.~(\ref{primitive equations}), ${\bf v}({\bf x},t)\! =\! (u,v)$
is the (eastward, northward) velocity in a frame rotating with
$\bf{\Omega}$, and $\Phi = gz$ is the geopotential, where $g$ is the
gravitational acceleration (assumed constant and includes the
contribution from the centrifugal acceleration) and $z$ is the height
above a fiducial surface (e.g., the 1 bar pressure level); ${\bf k}$
is the unit vector normal to the planetary surface; $f = 2 \Omega \sin
\phi$ is the Coriolis parameter, the projection of the planetary
vorticity vector $2{\bf\Omega}$ onto ${\bf k}$; $\nabla\!  _p$ is the
horizontal gradient on a constant-$p$ surface; $\omega \equiv Dp/Dt$
is the vertical velocity; $\alpha$ is the specific volume, the
reciprocal of density $\rho$; ${\cal F}$ and ${\cal D}$ represent
momentum sources and sinks, respectively; $\theta = T(p_{\rm
  ref}/p)^\kappa$ is the potential temperature\footnote{The potential
  temperature $\theta$ is related to the entropy $s$ by $ds = c_p
  d\ln\theta$.}, where $p_{\rm ref}$ is the reference pressure and
$\kappa = R/c_p$ with $R$ the specific gas constant; and,
$\dot{q}_{\rm net}$ is the {\it net} diabatic heating
rate.\footnote{Note that various heating ($ >\! 0$) and cooling ($ <\!
  0$) terms contribute to the diabatic heating rate $\dot{q}$---e.g.,
  stellar heating, longwave heating and cooling, latent heat, and
  sensible (conductive) heat.  Newtonian cooling approximation is one
  simple parameterization of $\dot{q}_{\rm net}$.} The above equations
are closed with the ideal gas law, $p = \rho RT$, as the equation of
state.

To arrive at Eq.~(\ref{primitive equations}), one begins with the full
Navier-Stokes equations---along with the energy equation, equation of
state, and boundary conditions---in the spherical geometry
\citep{Batchelor77}.  From the Navier-Stokes equations, two
approximations are made at the outset.  These are the ``shallow
atmosphere'' and the ``traditional'' approximations (cf.,
\citet{Salby96}).  The first assumes $z/R_p\ll 1$ and is valid for the
stable layers on all giant planets, including the extrasolar giant
planets.  The second is formally valid in the limit of strong
stratification, when the Prandtl ratio $(N/\Omega)^2\gg 1$.  In the
ratio,
\begin{equation}\label{bvfreq}
N(z)\ =\ \sqrt{\ \frac{g}{T}\, 
\left(\, \Gamma_{\rm ad} - \Gamma\, \right)\ }
\end{equation}
is the Brunt-V\"ais\"al\"a (buoyancy) frequency, where $\Gamma\!
_{\rm ad}\equiv g/c_p$ and $\Gamma\equiv -dT(z)/dz$ are the adiabatic
and the environmental lapse rates, respectively (Pedlosky 1987).  The
two approximations allow horizontal (perpendicular to {\bf k})
Coriolis terms to be dropped from the Navier-Stokes equations and
vertical accelerations to be assumed small, as will be further
described below.  As an example, consider \HDblah: if we assume
Jupiter-like composition, temperature range of $\sim$100 to
$\sim$1500~K across the globe, and nearly isothermal condition above
the tropopause\footnote{the level where the temperature ceases to
  decrease significantly with altitude (cf., \citet{Seager98})}, we
obtain $(N/\Omega)^2 \gsim 10^4$ with $N = g/(c_pT)^{1/2}$.  The
second approximation condition is also well met.  Hence, as in many
traditional atmospheric circulation studies, we may also reasonably
take Eq.~(\ref{primitive equations}) as our starting point for
extrasolar giant planets.
  
As can be seen in Eq.~(\ref{primitive equations}b), hydrostatic
balance is assumed in the primitive equations.  This condition
restricts the vertical motion to be slow compared to the horizontal
motion---or, equivalently, the vertical length scale of motions to be
small compared to the horizontal length scale.  It also allows local
temperature to be associated with the atmospheric mass.  That is,
given two pressure surfaces ($p_1 > p_2$) bounding a vertical region,
the local thickness (or mass per area) of the region, $\Delta z\equiv
z_2 - z_1$, can be related to its average temperature via the {\it
  hypsometric equation} \citep{Andrews87}:\\
\begin{equation}\label{hypso}
\Delta z\, =\, -H_p(\langle T\rangle)\, \ln\left(\frac{p_2}{p_1}\right), 
\end{equation}
where $H_p = R \langle T\rangle/g$ is the local pressure scale-height and\\
\begin{equation}
\langle T\rangle\, =\, \left[\,\int_{p_2}^{p_1} d\ln p\,\right]^{-1}\!
\int_{p_2}^{p_1} T({\bf x},t)\ d\ln p
\end{equation}
is the region's average temperature.  Note that $\Delta z$, $H_p$, and
$\langle T\rangle$ are all function of horizontal position and time.
Eq.~(\ref{hypso}) is the basis of the pressure coordinate.  Later, we
will use an analog of Eq.~(\ref{hypso}) to obtain horizontal
temperature distributions on extrasolar giant planets.

\subsection{Equivalent-Barotropic Formulation}
\label{sec:equivbarotrop}

For many applications, Eq.~(\ref{primitive equations}) is much too
unwieldy and broad in scope.  Moreover, in the absence of
observational information on the atmospheric vertical structure to
constrain the many parameters needed to initialize and solve
Eq.~(\ref{primitive equations}), a reduction of the equations is
warranted---and necessary (cf., \citet{Andrews87} and
\citet{Pedlosky87}).  A key element that must be represented
faithfully when modeling the global circulation and thermal structure
of planetary atmospheres is the dynamics of eddies and waves.  Eddies
and waves directly modify the flow and temperature structures on the
large-scale by transporting momentum, heat, and radiatively-active
species.  However, eddies and waves span in size from large scales
down to scales well below those resolvable by most models solving the
full primitive equations.  A complete series of high-resolution 3-D
calculations with all the key physics included and parameters explored
is currently computationally prohibitive.  Fortunately, some of this
difficulty can be alleviated.  In this work, we accomplish the
reduction of the equations, resolving of the eddies and waves, and
exploration of the parameter-space by vertically integrating
Eq.~(\ref{primitive equations}) and using the well-tested and
highly-accurate pseudospectral method \citep{Orszag70,Eliassen70} to
solve the resulting, simpler set of equations.  In doing so we are
able to focus on some of the important elements critical for
accurately capturing the global dynamics.

There are other justifications for focusing on the column-integrated,
horizontal dynamics as well.  First, the region of extrasolar giant
planet atmosphere that requires modeling for current observations is
an extremely thin layer within the overall stratified layer, at most a
few scale heights thick and located near the tropopause.  It is in
this sub-layer where most of the stellar irradiation is seen to be
absorbed in radiative transfer calculations
\citep[e.g.][]{Seager05,Iro05} and from where the strongest (longwave)
sensible features are likely to emerge observationally.  In addition,
given the nearly isothermal vertical structure in that region
\citep[e.g.][]{Seager98,Sudar00}, we expect the flow to be vertically
well aligned.\footnote{Actually, good alignment may exist in the
  entirety of the stratified layer since the temperature structure is
  nearly isothermal throughout the layer.}  In atmospheric dynamics
studies, the use of reduced vertical-resolution models is
common---especially for the tropopause region and above, where there
is a strong barotropic component (i.e. vertical alignment; cf.,
\citet{Juckes87, Salby90, Polvani95, Ferreira97}).  The typical length
scale of the most unstable baroclinic mode (related to the
non-alignment) is the Rossby deformation radius, \LR~$= NH/f\approx
(gH)^{1/2}/f$, where $H\!\sim\!  H_p$ is the characteristic thickness
\citep{Pedlosky87}.  For baroclinic waves to exert a strong presence,
\LR\ should neither be $\ll\! R_p$ nor $\gg\! R_p$.  As noted, the
extreme smallness of \LR\ has already been exploited to great
advantage in modeling Solar System giant planet circulations with a
barotropic model \citep{Cho96b}.  For most unsynchronized extrasolar
giant planets, we expect similar, or only slightly larger, values of
\LR\ ($\ll\! R_p$).  Even for close-in extrasolar giant planets, which
receive intense heating, there is some justification for using a
barotropic model since \LR\ values for them are generally $\!\gsim
R_p$ \citep{Cho03,Menou03}. In the future, it will be important to
{carefully} delineate the regime of validity of our barotropic
models with baroclinic (i.e., multi-layer) extensions.

\subsubsection{Equations}

Remarkably, \citet{Salby89} has shown that both adiabatic
($\dot{q}_{\rm net} = 0$) and diabatic ($\dot{q}_{net}\ne 0$) forms of
Eq.~(\ref{primitive equations}) directly reduce to a set of equations
similar to the shallow-water equations, a much studied set of
equations in geophysical fluid dynamics.  The shallow-water equations
are the non-linear version of the Laplace tidal equations
\citep{Laplace78,Gill82}.  Because these dynamical equations reveal
the clearest interpretation of the role of heating in isentropic
coordinates, we have used that coordinate form of the equations in
this work.  In isentropic coordinates, constant $\theta$-surfaces form
the vertical coordinates, rather than $p$- or $z$-surfaces.  The new,
reduced set of equations---the equivalent-barotropic equations in
$\theta$-coordinates---are obtained by assuming that the constant
surfaces of thermodynamic variables ($p$, $\rho$, $T$, $\theta$) share
a common horizontal structure.  A salient feature of the
equivalent-barotropic model, as well as the shallow-water model, is
that it is a global atmospheric dynamics model that can include the
combined effects of differential rotation, stratification, and
small-scale eddies and waves over long duration integrations.  It also
reproduces qualitative features of the global circulation on giant
planets with a minimal set of adjustable parameters and assumptions.
In this paper, we present results for the adiabatic case only, in
which the effect of heating is parameterized by deflecting the modeled
surface consistently with the temperature at the surface (see below
and \S3.3).  A forthcoming companion paper will discuss the diabatic
case.

The equivalent-barotropic equations govern the dynamics of a
semi-infinite gas layer, which is bounded below by a material surface.
The bounding surface deforms according to the local temperature on the
surface.  In this work, the material surface is an isentropic surface,
$\theta_0 =$~constant.  The governing equations read \citep{Salby89}:
\begin{mathletters}\label{equivalent-barotropic equations}
\begin{eqnarray}
\frac{D {\bf v\ }_{}}{D t\ }  & = & -\, 
{\bf \nabla} \left[\,{\cal H} + {\cal H}_B\,\right]\ -\ 
f {\bf k \times v}\ +\ {\cal D}_{\bf v}\ , \\
\frac{D {\cal H}\ }{D t\ }  & = &
-\kappa{\cal H}\,\nabla\!\cdot\!{\bf v}\ +\ {\cal D}_{\cal H}\ ,
\end{eqnarray}
\mbox{where}
\begin{equation}
\frac{D}{D t} \ = \ \frac{\del}{\del t} + 
{\bf v}\!\cdot\!\nabla\, ,
\end{equation}
\begin{equation}
{\cal H}\ =\ \frac{\theta_0}{{\cal A}_0\Gamma}\left(\frac{p_0}{p_{\rm
  ref}}\right)^\kappa\, ,
\end{equation}
and
\begin{equation}
{\cal H}_B\ =\ \frac{z_0}{{\cal A}_0}\, .
\end{equation}\\
\end{mathletters} 
The boundary condition,
\begin{equation}\label{boundary_condition}
\frac{D\theta_0}{Dt\, } = 0\, ,
\end{equation}
is trivially satisfied in the adiabatic case.  In the diabatic case,
this equation with forcing terms becomes a formal addition to the set
of equivalent-barotropic equations.  In
Eq.~(\ref{equivalent-barotropic equations}), the dependent variables
are the barotropic transformations of the original baroclinic
variables in Eq.~(\ref{primitive equations}).  For example,
\begin{equation}\label{structf}
{\bf v}({\bf x},t) = -\frac{1}{p_0}\,
\int^0_{p_0({\bf x},t)} {\bf v}({\bf x},p, t)\ dp\, ,
\end{equation}\\
for the baroclinic velocity, ${\bf v}({\bf x},p,t)$.  Other variable
have been similarly transformed.  An equivalent-barotropic structure
function, ${\cal A} = {\cal A}(\theta(p))$, is defined
such that the inverse transformation for ${\bf v}$ is\\
\begin{mathletters}
\begin{equation}\label{separ}
{\bf v}({\bf x},p,t)\ =\ {\cal A}(\theta)\, {\bf v}({\bf x},t)\, ,
\end{equation}
with the normalization,
\begin{equation}
\frac{1}{p_0}\,
\int^{p_0({\bf x},t)}_0 \ {\cal A}\ dp\ =\ 1\, .
\end{equation}\\  
\end{mathletters}
In Eqs.~(\ref{structf}) and (\ref{separ}), barotropic and baroclinic
forms are both represented by the same variable, but that should not
elicit confusion.  The ${\cal D}$'s in
Eqs.~(\ref{equivalent-barotropic equations}) represent dissipation, as
before.  The subscript ``0'' (as in $p_0$ or $\theta_0$, for example)
refers to the value of the variable at the bottom bounding surface.
Accordingly, ${\cal A}_0 = {\cal A}(\theta_0)$ is a measure of the
local baroclinicity (i.e., vertical shear, or lack of vertical
alignment), at the bounding surface.  For an equivalent-barotropic
structure which decays (grows) vertically, ${\cal A}_0$ must be
greater (less) than 1 to preserve the normalization condition.  In
strongly-stratified atmospheric regions, in which the flow structures
tend to be vertically aligned, ${\cal A}_0$ should be close to unity.
We have varied ${\cal A}_0$ from 0.8 to 1.2 and find no qualitative
changes in the results.\footnote{{In general, the results are
sensitive to large differences in ${\cal A}_0$.}} In all the
simulations presented in this paper, ${\cal A}_0 = 1$.  In
Eq.~(\ref{equivalent-barotropic equations}e), $z_0 = z_0({\bf x},t)$
is the prescribed elevation of the bounding surface.  Using the
definition of $\theta$, we have:
\begin{equation}\label{thickness}
{\cal H}\  =\  \frac{T_0}{{\cal A}_0\,\Gamma} = 
\frac{H_{p0}}{{\cal A}_0\ \kappa}\, .
\end{equation}
Thus, the thickness, ${\cal H} = {\cal H}({\bf x},t)$, is essentially
the temperature at the bounding surface.  It is closely related to the
local pressure scale height evaluated at the bounding surface, $H_{p0}
= RT_0/g$, and defines the potential temperature scale height,
$H_{\theta 0} = H_{p0}/\kappa$.  It can be readily seen from
Eqs.~(\ref{equivalent-barotropic equations}) that prescribed heating
forces the flow through the deflection of the bounding surface, which
advects the temperature.  The advected temperature in turn further
drives the flow, {when gradients form}.

When $\kappa\! =\! 1$, the set of Eqs.~(\ref{equivalent-barotropic
equations}) is formally identical to the shallow-water equations, with
bottom topography (cf., \citet{Pedlosky87}).  From the point of view
of shallow-water dynamics, $\kappa$ (which is always $< 1$) primarily
produces enhanced advection.  As in the shallow-water equations,
Eqs.~(\ref{equivalent-barotropic equations}) admit an important
conservation law for the potential vorticity $q$:\\
\begin{mathletters}\label{PVD}
\begin{equation}
  \frac{Dq}{Dt} = {\cal D}_q \, ,
\end{equation}
where
\begin{equation}
 q\, ({\bf x},t)\equiv
\left[\,\frac{\ \zeta\ +\ f\ }{{\cal H}^{1/\kappa}}\,\right]
\end{equation}
\end{mathletters}\\
and ${\cal D}_q = {\cal D}_q({\cal D}_{\bf v},{\cal D}_{\cal H})$ is
potential vorticity dissipation.  In the absence of dissipation, the
conservation law becomes:\\
\begin{equation}\label{PVC}
\frac{Dq}{Dt} = 0\, .
\end{equation}
Eqs.~(\ref{PVD}) and (\ref{PVC}) are the equivalent-barotropic
generalization of the shallow-water potential vorticity conservation.
According to Eq.~(\ref{PVC}), $q$ is materially conserved in the
absence of dissipation and can therefore serve as a proper tracer of
the flow.  In this work we will make frequent use of $q$ to visualize
the flow.  In observed atmospheres, $q$ is conserved very well,
especially near the tropopause and above.  Isentropic maps of $q$ have
been used very effectively to gain much understanding in atmospheric
dynamics studies of the Earth (cf., \citet{Hoskins85}).  Indeed, one
of the significant advantages of our approach is our ability to
conserve this crucial quantity to a high degree.  This is achieved
through the high resolution of our calculations and the accuracy of
the employed numerical method, which we now briefly describe.

\subsubsection{Numerical Solutions}

Eqs.~(\ref{equivalent-barotropic equations}) are solved numerically in
full spherical geometry---with ${\bf x}\! =\! (\lambda,\phi)$, where
$\lambda$ is the longitude and $\phi$ is the latitude.  The equations
are solved in the vorticity-divergence form---i.e., the curl and
divergence of the Eqs.~(\ref{equivalent-barotropic equations}a), along
with Eq.~(\ref{equivalent-barotropic equations}b).  In this form, the
field variables, ($\zeta$, $\delta$, ${\cal H}$) =
($\bf{k}\!\cdot\!\nabla\!\times\!\bf{v}$, $\nabla\!\cdot\!\bf{v}$,
${\cal H}(T_0)$), are projected onto the Legendre basis via the
Gauss-Legendre transform \citep{Orszag70,Eliassen70}:
\begin{equation}\label{basis}
\xi (\lambda,\phi, t)\ =\ \sum_{n=1}^{N=N(M)}\!\sum_{m = -n}^n\  
\hat{\xi}_n^m\, P_n^m(\phi)\, e^{im\lambda}\ ,
\end{equation}
where $\xi$ represents an arbitrary field variable, $P_n^m$ is the
associated Legendre polynomial for total wavenumber $n$ and sectoral
wavenumber $m$, and $N$ and $M$ are the total and sectoral
truncations.  When $N = M$, the truncation mask is a triangle in
spectral space, and the resolution of a calculation is referred to as
``T106'', for example, if $N=106$.  In physical space, this truncation
corresponds to a minimum of (longitude, latitude) = ($3N\! +\! 2 =
320$, $(3N\!  +\!  2)/2 = 160$) grid points \citep{Orszag70}.  In most
cases, a spectral calculation is computationally much more accurate
(though less efficient), per degree of freedom, than a finite
difference calculation \citep{Canuto88,Haltiner80}.  The high
resolution and accuracy achieved in the spectral algorithm is crucial
for allowing small-scale structures and turbulence to evolve without
suffering much dissipation.  A small hyperviscosity, of the form
$(-1)^{\chi+1}\nu\nabla^{2\chi}$ for $\chi = \{1,2,3,4,8\}$, is
included in Eq.~(\ref{equivalent-barotropic equations}) to stabilize
the numerical integration while extending the turbulent inertial
range.  Details on the use and effects of hyperviscosity can be found
in \citet{Cho96a}.  As for the time-integration, the second-order
accurate leapfrog scheme is used.  We have thoroughly explored the
numerical parameter-space associated with the present work and
identified the region in which our results do not depend qualitatively
on the precise choice of numerical parameter values.

Throughout this paper, simulations described are initialized with a
random turbulent flow.  This initialization crudely represents the
effect of stirring arising from a variety of barotropic and baroclinic
processes; its use is nearly universal in turbulence simulation
studies.  The stirring is applied only at the start of the simulation
and does not represent a continuous source of small scale forcing: it
is used to generate a reasonably constrained background flow in our
atmospheric simulations.  The motivation behind this key ingredient of
our modeling strategy is further discussed in \S3.4.  The random
vorticity field is characterized by a narrow band in spectral space
with a specified mean amplitude that corresponds to a global root mean
square velocity $\bar{U}$.  The divergence field is set to be
uniformly zero initially.  Given these fields, the remaining variable,
${\cal H}$, is obtained by requiring that the initial flow be
nonlinearly balanced---i.e., satisfy the divergence of
Eq.~(\ref{equivalent-barotropic equations}b) in the limit $\del / \del
t \rightarrow 0$.  This procedure allows the turbulent calculation to
proceed over long times without ``blowing up''.  A detailed discussion
of the initialization procedure is also given in \citet{Cho96a}.  We
have verified that the results are insensitive to the location of the
initial distribution's peak in spectral space, as long as the location
is much smaller than the scale defined by $L_{\beta}$ (Rhines scale).
The mean amplitude of the initial perturbations is a free parameter of
the model, which we express here in terms of $\bar{U}$.

{The equivalent-barotropic model admits the following parameters:
${\cal A}_0$, $\kappa$, $R$, $\Omega$, $g$, $R_p$, $\bar U$, and
$\bar{T}_0$.  Here, $\bar{T}_0$ is the global mean $T_0$.  Note that
three parameters appear in the equivalent-barotropic model, in
addition to the ones that are present in the shallow-water model.
They arise from the more explicit representations of the physics in
the equivalent-barotropic model, compared to the shallow-water model.
The added cost of the extra variables is minimal, however, since
nearly all of the needed parameters can be reasonably constrained by
observations or plausible scaling analysis.}

{In the present work, if Jupiter is used as a paradigm giant planet to
constrain $R$ and $\Omega$ (for the case of unsynchronized extrasolar
giant planets), $\bar{U}$ becomes the most uncertain of the
parameters.  For Jupiter, $R$ is 3779~J~Kg$^{-1}$~K$^{-1}$, using
$\kappa\! =\! 2/7$; we have found that a $\kappa$ value of 2/5,
appropriate for monatomic gases, also does not qualitatively change
the results.  In addition, $\bar{U}\approx70$~m~s$^{-1}$ for Jupiter;
note that $\bar{U}$ in the upper tropospheres of all the Solar System
planets is known and is between $\sim$30~m~s$^{-1}$ and
$\sim$400~m~s$^{-1}$.  However, the appropriate value for current
extrasolar giant planets is not known and not easily constrained,
leading us to adopt a somewhat-uncomfortable wide range:
$\sim$10$^2$~m~s$^{-1}$ to $\sim$10$^3$~m~s$^{-1}$.  While not
stultifying, the uncertainty is not insignificant.  As we shall show,
values in the lower range can lead to identifiably different
circulation pattern than those in the upper range.  As yet, we are
unable to definitively conclude on the exact circulation pattern
without further observational constraints.  Currently, there is no way
to obtain $\bar{U}$ from first principles.}

\subsection{Thermal Forcing}
\label{sec:forcing}
Two points are important to note when addressing thermal forcing on
extrasolar giant planets: {\it i}) in general, it is {\it not} the
direct stellar irradiation but the net flux (i.e. that which is
non-linearly modified by the circulation) that forces the temperature
tendency (Eq.~(\ref{primitive equations}d)) and balances the flow;
and, {\it ii}) because some of the stellar irradiation will be
rejected by the atmospheric gases and clouds, the radiative
equilibrium temperature distribution is not known with certainty for
extrasolar giant planets at the present time.  Further observations
are required.  Note that {\it ii}) necessarily implies that the
strength and distribution of forcing in {\it i}) is unknown for
extrasolar giant planets and must be treated as a parameter.

For the Solar System giant planets, for which good flux observations
exist, it is reasonable to model their atmospheres as layers of
uniform thickness (temperature).  This is because the emergent flux
(which is as much as, or larger than, the insolation) is nearly
uniform on them and $\tau_r\!\gg\!\tau_a$.  Indeed, the
flux-associated equator-pole temperature gradient on Jupiter is only
$\sim$3~K at all longitudes.  This suggests that thermodynamics may be
effectively decoupled from the dynamics, as has been done in
\citet{Cho96b}.  Such a prescription is also probably appropriate for
many unsynchronized extrasolar giant planets, which may behave
plausibly like a ``warm Jupiter''.  However, the uniform thickness
assumption is not likely to be as valid for the extremely close-in
giant planets, given the much smaller $\tau_r$ than for Jupiter and
the expected high radiative--equilibrium temperature gradients.  On
these planets, the {true} temperature at the substellar point should
be closer to the equilibrium temperature than in the polar---and if
synchronized, antistellar---regions.  {Hence, an estimate of the
radiative--equilibrium temperature distribution is required to model
the flow on close-in giant planets.}

For synchronized planets, in the absence of circulation, the radiative
equilibrium temperature on the day side is modeled as
\begin{equation}
\label{eq:localradeq}
T_{\rm eq}(\lambda,\phi)\ =\ T_* \left( \frac{R_*}{D} \right)^{1/2}
\left( 1-A_{\lambda\phi} \right)^{1/4} \left( \cos\lambda\,\cos\phi
\right)^{1/4},\\
\end{equation}
where $T_*$ and $R_*$ are the parent star's effective temperature and
radius, respectively.  Here, $D$ is the distance of the planet from
the parent star; and, $A_{\lambda\phi}$ is the local value of the
albedo, a function of the location on the day side.  This definition
differs from the standard, ``averaged'' one, often found in the
literature:
\begin{equation}
\label{eq:globalradeq}
\bar T_{\rm eq}\ =\ T_* \left( \frac{R_*}{2D} \right)^{1/2} 
\left[\, \hat{f}\,(1-A)\,\right]^{1/4}, \\
\end{equation}
where $A$ is the geometric albedo (an averaged value over the entire
day side) and $\hat{f}$ is a parameter that describes whether the
stellar flux is re-emitted by the entire planet ($=\! 1$) or only the
day side~($=\! 2$).  Note that because the albedo is a function of
temperature and space (via cloud formation, for example), the
radiative equilibrium temperature field itself is not really well
defined---even in the absence of motion.  However, these two
definitions allows us to very roughly bracket the range of forcing
amplitude that needs to be explored.

Prior to synchronization, the outgoing flux was more likely
``homogeneously distributed'' over the planet's surface.  There is no
orography or topography to make the diabatic heating zonally
asymmetric on a gaseous giant planet.  The temperature resulting from
a planet which emits homogeneously is given by $\bar T_{\rm eq}$ in
Eq.~(\ref{eq:globalradeq}) with $\hat{f}\! =\!  1$.  After reaching
synchronization, however, the incoming stellar flux distribution is
not zonally-symmetric and the equilibrium temperature associated with
this influx is not uniform.  The maximum radiative equilibrium
temperature is then given by $T_{\rm eq}\,(\lambda\!  =\!0,\phi\!  =\!
0)$ in Eq.~(\ref{eq:localradeq}) at the substellar point with the
temperature profile decreasing as $(\cos\lambda\, \cos\phi)^{1/4}$
away from the point.  Statically, the thermal forcing resulting from
synchronization can therefore be modeled in first approximation as a
permanent temperature perturbation superimposed on a layer of nearly
uniform temperature, $\bar T_{\rm eq}$.  If the differences in local
vs.\ global albedo values are ignored (i.e., $A = A_{\lambda\phi}$
assumed) for the sake of argument, Eqs.~(\ref{eq:localradeq})
and~(\ref{eq:globalradeq}) indicate that the amplitude of thermal
forcing at the substellar point could be as large as $\sim\! 40$\% in
excess of the uniform temperature value, $\bar T_{\rm eq}$ (obtained
for $\hat{f}\! =\!  1$).  Given the above-mentioned uncertainties in
the definition of radiative equilibrium itself, we will consider
amplitudes of thermal forcing reaching up to 20--40\% in our
circulation models.

On the night side, absent any circulation, atmospheric temperature
should be approximately the effective temperature determined by the
internal luminosity.  The internal luminosity is powered by the slow
gravitational contraction and possible helium rainout in the
planet\footnote{Other mechanisms may be at play as well (cf.,
  \citet{Guillot06}).}:
\begin{equation}
T_{\rm eff}\ \sim\ 100\, {\rm K}\ 
\left( \frac{L_{\rm int}}{L_{\rm int,Jup}}\right)^{1/4} 
\left( \frac{R_p}{10^{8}~{\rm m}}\right)^{-1/2}\, ,
\end{equation}
where $L_{\rm int,Jup}\sim 8.7\times 10^{-10}$~J~s$^{-1}$ is the
estimated internal luminosity of Jupiter.  With the observed
temperature at the substellar point of \HDblah\ possibly as high as
$\sim$1700~K (cf., \citet{Deming05,Seager05}), the temperature
difference between the substellar and antistellar points on this
extrasolar giant planet could be as high as $\sim$1600~K in the
absence of motion.  Obviously such an enormous temperature difference
would be quickly, and violently, neutralized by atmospheric motion as
long as $\tau_a < \tau_r$.  It is crucial to note that, if $\Omega\ne
0$, the readjusted flow state will {\it not} be that which merely
balances the above temperature gradient but also one which is very
strongly modified by the planetary rotation.  This is a fundamental
property of rotating fluids \citep{Gill82}.  Note also that $\tau_r$
is locally a nonlinear function of $T$ and of the motion.  Venus,
which has a remarkably homogeneous temperature distribution despite
its extremely slow longitudinal variation of insolation, may serve as
an example of such a {\it post}-adjusted distribution.  Incidentally,
during such an readjustment phase, the flow state may not be unlike
the turbulent initial condition used in our simulations, since
small-scale gravity waves and turbulence will be generated during the
process.

The salient point here is that a reasonable response to irradiation is
to make the atmospheric layer thicker (due to the warmer radiative
equilibrium temperature) on the day side than on the night side, as
dictated by Eq.~(\ref{hypso}). Accordingly, in this work, the flow is
forced with heating by appropriately ``puffing up'' the atmospheric
layer on the heated side while maintaining a global equilibrium
temperature $\bar{T}_0$ (=~$\bar{T}_{\rm eq}$).  This is accomplished
by introducing a permanent deflection of the bounding material surface
through a specified $z_0({\bf x})$ in Eq.~(\ref{equivalent-barotropic
  equations}e).  In most runs, the deflection is ``grown'' after the
basic flow pattern is established with a characteristic $e$-folding
time (usually $\sim$5$\tau$, where $\tau\!\equiv\! 2\pi/\Omega$ is the
rotation period) in order to minimize introducing a large unbalanced
component (small-scale fast modes) in the flow.  A sudden
``turning-on'' of the forcing introduces a large amount of fast modes,
which are in general insignificant for meteorology and are filtered out
in numerical weather predictions \citep{Haltiner80}.  No heat
(thickness) is added overall, but merely redistributed by the flow.
Generally, this is most appropriate when $\tau_r\gg\tau_a$ or when
$\dot{q}_{\rm net} \to 0$ (Eq.~(\ref{primitive equations}d)).  Here,
we are mainly concerned with the horizontally advected component of
the temperature, neglecting exchanges with regions above and below the
modeled layer.  However, as we argue in \S3.4, because of the way in
which the bounding surface is deflected, our steady forcing procedure
may also capture important features of the opposite regime, in the
fast radiative limit $\tau_r \to 0$.

The fluid layer is bulged on the day side according to:
\begin{mathletters}\label{forcing}
\begin{equation}
 {\cal H}(\lambda,\phi)\ =\ {\bar{\cal H}}\ +\ 
{\cal H}^\prime \cos\phi\cos\lambda
\end{equation}
and
\begin{equation}
 z_0(\lambda,\phi)\ =\ {\bar z_0}\, -\, z_0^\prime
 \cos\phi\cos\lambda\, ,
\end{equation}
\end{mathletters}\noindent
where ${\cal H}^\prime,\, z_0^\prime > 0$ are the constant, maximum
perturbation amplitudes of forcing at the substellar point above the
reference levels, $\bar{\cal H}$ and $\bar{z_0}$.  ${\cal H}^\prime$
is set at the start of the simulation to ensure that the flow
is down-gradient of the temperature distribution and then to 0
  thereafter.  Without loss of generality, we set $\bar{z_0} = 0$,
since it can be subsumed into $\bar{\cal H}$, but $z_0^\prime$
  is held fixed throughout the simulation.  That is, a permanent
  deflection of the lower bounding surface is maintained while a mass
  flow initially away from the substellar point is ensured, consistent
  with heating.  The same perturbation amplitude, but with opposite
sign, is applied to the layer on the night side so that the forcing
integrated over the entire planet surface $S$ is zero and
\begin{equation}
\frac{1}{4\pi}{\int_S\ {\cal H}(\lambda,\phi)\ d\phi\,d\lambda}\ =\
\bar{\cal H}\, .
\end{equation}
We have also performed simulations in which the day side merges
smoothly into a flat night side and found that the results do not
change qualitatively.  According to the argument outlined above, if we
define
\begin{equation}
\eta\ \equiv\ \frac{\ {\cal H}^\prime\ +\ z_0^\prime\ }
{{\bar{\cal H}}}\ , 
\end{equation}
we are interested in $\eta\le 0.4$.  Since the actual value of $\eta$
may be different for different extrasolar giant planets and different
altitudes, we vary its value from $0$ to $0.4$ in our simulations.

Note that the deflection approach cannot be used in the much lower
levels of extrasolar giant planets since the vertical entropy gradient
becomes very shallow in the equatorial region and a $\theta$-surface
cannot be practically defined there.\footnote{Computationally, the
large deflection required causes holes to develop in the model layer,
leading to blow-up.}  The bulging associated with heating can in fact
be represented either indirectly by lowering the height of the lower
bounding surface (thus producing a ``surface deficit'') or directly by
adding mass (thus increasing $\cal H$).  Similarly, the reduced
thickness of the fluid layer on the night side (due to cooling) can be
represented by directly removing mass, or by increasing the height of
the lower bounding surface.  This is just a restatement of the
hypsometric relation, Eq.~(\ref{hypso}).  The direct method is
``diabatic'' and the indirect method is ``adiabatic''.  We have
investigated both methods but present only the adiabatic case in this
paper.  Extensive study of the diabatic case will be presented
elsewhere.  Our simulations also indicate that the key results on the
developed circulation do not depend qualitatively on the precise
functional form of the differential thermal forcing---e.g., Gaussian
or ``$\cos^{1/4}$'' dependence, as opposed to the simple ``cosine''
dependence of Eq.~(\ref{forcing}).
  
Because the bulge associated with thermal heating can introduce a
large unbalanced component in the flow, as described above, it can
quickly lead to a numerical instability.  Recall that the pre-existing
part of the flow is nonlinearly balanced.  Unless the flow is
``stabilized'' with draconian dissipative measures, which are
unphysical, a gentle ``ramp up'' to the equilibrium profile is
necessary.  Even in this case, the destabilizing effect of imbalance
when the amplitude is large cannot be completely avoided.  We have
tested both instantaneous and gradual forcing schemes. In the
instantaneous version, the fluid is fully bulged right at the
beginning of the simulation. In the gradual version, forcing is slowly
increased with time during the numerical simulation, up to the chosen
$\eta$ value.  Again, we have found that simulations with
instantaneous or gradual forcings gave results that are qualitatively
similar.  Overall, we find that the developed flow is surprisingly
robust under a wide variety of conditions.  This is one of our
principle conclusions from this study.

\subsection{Key Assumptions and Potential Caveats}

{At this point, it is useful to review the main assumptions in our
modeling strategy, in order to remind the reader of some of the
potential limitations of the results presented in the ensuing
sections.  We model atmospheric circulation on close-in extrasolar
giant planets with a single, adiabatic, equivalent-barotropic layer,
which is steadily forced on the large scale as described in \S3.3.
Except in specific cases, a prescribed amount of small-scale
turbulence is applied one time at the start of the model simulation.
This is to generate a background flow which is consistent with one
under no thermal forcing and which interacts with one forming due to
our steady thermal forcing.  Our aim here is to study, in a highly
idealized way, the interaction between the background and forced
flows.}

{Focusing on equivalent-barotropic flows allows high enough spatial
resolution to simultaneously address the important flow interaction
and explore the parameter-space well enough to ensure robustness of
the results.  While we have provided some justifications for the
relevance of equivalent-barotropic character of the modeled region of
the atmosphere, baroclinicity due to vertical variations is
significant and must be included for additional physical realism.
Hence, it is important to extend the present one-layer calculations to
multi-layer calculations.  Ideally, high resolution should also be
employed in the vertical direction for the same reason given for the
horizontal direction.  Extra care should be given to the specification
and effects of vertical boundary conditions as well.}

{The assumption of adiabaticity is clearly a simplification and one
that is not applicable to all regions of the atmosphere.  However, it
is valid in many regions, where pressure is high, temperature is low,
or heating and cooling cancel in the net.  In addition, it allows the
effects of diabatic forcing to be put in lucid context.  In
particular, the present study is a first step towards assessing the
crucial interplay between ``adiabatic wind'' (as parameterized by
$\bar U$) and ``diabatic wind'' (as parameterized by $\eta$).  This
step is significant because the relationship between wind structure
and thermal forcing on planetary atmospheres is complicated and not
completely understood: currently, there is no theory that explains the
magnitude of the observed wind speeds at cloudtop heights.  Hence, the
wind speed cannot be naively related to the amount of incident stellar
flux.  Table~\ref{tab:one}, showing a non-trivial scaling between
$\bar U$ and semi-major axis $a$ for Solar System giant planets,
exemplifies this point.  In this respect, atmospheric simulations
which attempt to predict wind speeds purely on the basis of applied
thermal forcing are physically limited, even though such model setups
may be conceptually appealing.}

{One regime of circulation for which our adiabatic simulations may
turn out to be useful is the short radiative cooling time limit,
$\tau_r \to 0$.  This may appear counterintuitive at first since the
usual assumption of adiabatic condition corresponds to the opposite
limit, $\tau_r \to \infty$.  Notice, however, that in the limit
$\tau_r \to 0$, a parcel of air relaxing according to Newtonian
``cooling'' adjusts instantaneously to the state of radiative
equilibrium (unless the ``cooling'' is balanced by other effects, such
as conduction). If $\tau_r \ll \tau_s$, the sound crossing timesscale
($\equiv\! R_p/c_s$), such a fast forcing of the thermodynamic
equation (Eq.~[\ref{primitive equations}d]) cannot be balanced by
global-scale motions.  In this limit, one might sensibly filter the
fast process out of the dynamics, similar to the way in which sound
waves are filtered out by imposing hydrostatic balance or gravity
waves are filtered out by imposing quasi-geostrophy.  Note that when
$\tau_r\! \sim\! \tau_a$, the adiabatic approach is not justified and
explicit diabatic calculations are necessary.}

{Another major assumption in our modeling approach is that extrasolar
planet atmospheres are turbulent, in the sense that the flow is
dynamically active on many scales, and that turbulent eddies and waves
play a significant role in establishing the observed flow.
Turbulence, as signified by the presence of eddies and waves, in the
atmosphere is produced by a variety of mechanisms; and, it is present
on all Solar System planets---even if visible tracers like clouds are
not there to spotlight it.  In our simulation set-up, one may question
the validity of using a small-scale turbulent forcing for close-in
extrasolar giant planets: given the expected dominance of large-scale
external irradiation, one might argue for the absence of small-scale
stirring---particularly from obvious sources such as convection and
baroclinic instability.  Here, we emphasize that the small-scale
forcing in our simulations is not used to literally model those
processes but used to facilitate a dynamically constrained atmosphere
that is not at rest.  After the initial one-time stirring, the flow in
our close-in extrasolar planet simulations is forced only with a
steady, large-scale forcing.}

{In summary, we recognize the difficulty of assessing the
applicability of our assumptions on extrasolar giant planetary
atmospheres in the absence of almost any direct observational
constraints.  Since key ingredients necessary to reliably model the
atmospheric dynamics of close-in extrasolar giant planets are
presently not determined, we have adopted an approach that emphasizes
the following: 1)~previous successful modeling strategy, 2) physical
processes that are reasonably well understood on Solar System planets,
and 3) a baseline that allows effects of additional physics, such as
diabatic heating, to be clearly delineated.  It is standard to employ
such an approach initially in atmospheric studies.}

\section{Unsynchronized Giant Planets}

\subsection{Solar System Giant Planets} \label{sec:ssebf}

Given the limited amount of direct information on atmospheres for
extrasolar giant planets, validation of theoretical models is
critical.  The current work is closely-related to the shallow-water
equations studies by \citet{Cho96a,Cho96b}.  Their findings are
directly relevant and lends some credence to our work.  Hence, we
briefly summarize the salient features before presenting results with
the equivalent-barotropic equations.  

For shallow-water equations in spherical geometry, two dimensionless
numbers define the dynamics for a fixed radius: the Rossby number, $
R_o~\equiv~U/(fL)$, and the Froude number, $F_r\equiv U/(gH)^{1/2}$.
These numbers arise from the ratio of terms in the momentum equation
of the shallow-water equations.  The numbers also appear in the
equivalent-barotropic equations and control the dynamics in the same
way.  $R_o$ is a measure of relative vorticity compared to the
planetary rotation (or, the rotation time compared to the advection
time), and $F_r$ is the shallow-water analog of the Mach number in
compressible flows.\footnote{In the shallow-water equations, surface
  gravity waves are admitted, not sound waves, since the fluid is
  assumed homogeneous and incompressible in 3-D.}  A convenient and
important third dimensionless number can be constructed from the above
two: the Burger number, $B_u\equiv R_o^2/F_r^2 =$ \LR$^2/L^2$.  In
rotating-stratified fluids, \LR\ acts as an $e$-folding length
limiting vortex and jet interactions.  In planetary applications, the
$\beta$ (Rhines scale) effect is necessary, but not sufficient, to
produce long-lasting banded structures; \LR\ must also be finite.
That is, the fluid must have a free surface, or be 2D-compressible.
Quantitatively, \LR\ must be $\!\lsim R_p/3$ for a banded structure to
be stable.  In their studies, \citet{Cho96a,Cho96b} use the simplest
form of shallow-water equations, without external forcing or
topography, in order to clearly delineate the intrinsic flow evolution
from that due to forcing.  Only a delta-function force in time is used
to stir the fluid at the beginning of the simulation.  In the cases in
which a simple, continuous Markovian random forcing is used, the
physical-space evolution is strongly influenced by the forcing and
vortices do not merge robustly \citep{Cho97}.

Because of the close relationship, shallow-water equations results can
be used to validate the current model.  We do this by performing
simulations of Solar System giant planet atmospheres using the
adiabatic equivalent-barotropic equations.  This not only gives
confidence in the present code, it also sets the context for more
complex diabatic calculations.  In addition, direct comparisons with
shallow-water equations simulations allow the effects of additional
physical parameters in equivalent-barotropic equations simulations to
be carefully assessed.  Since the primitive equations require even
greater number of parameters, rendering model results sometimes
difficult to interpret clearly, equivalent-barotropic equations
simulations provide a good benchmark for full primitive equations
models.  The use of shallow-water equations model to elucidate or
verify primitive equations model results is a common procedure in
atmospheric dynamics studies.  Moreover, present calculations of Solar
System giant planets are useful in themselves since they have never
been simulated with the equivalent-barotropic equations.  The
calculations also have value in that Solar System giant planets still
serve as paradigmatic giant planets in many ways; there must surely be
extrasolar giant planets which are Solar System giant planet-like in
their atmospheric properties.  Finally, we note that the
equivalent-barotropic equations are preferable to the shallow-water
equations because they derive directly from the primitive equations;
certain parameters (e.g., \LR) acquire a more physical, less
arbitrary, interpretation than in the shallow-water equations.
 
Figures~1--\ref{ss_jets} show examples of Solar System giant planet
simulations from our study.  The global planetary parameter values
adopted for the simulations are listed in Table~\ref{tab:one}.  The
values for \HDblah\ are also included in the Table for comparison.  In
the simulations, small finite eccentricities are ignored.  Since it is
known from observations that the emitted flux is fairly uniform over
the entire surface of all Solar System giant planets, we do not
include any differential thermal forcing in our simulations of these
planets.  The qualitative success of these calculations in reproducing
the global circulation pattern of Solar System giant planets gives
confidence in our approach to modeling these planets and
unsynchronized extrasolar giant planets with small eccentricities.

Figures~\ref{jupiter_side} and \ref{jupiter_top} show a typical
evolution from one of our Jupiter simulations (Model~S1 in
Table~\ref{tab:two}).  Contour maps of the potential vorticity ($q$)
field are presented.  Recall that $q$ is a tracer of the fluid, hence
a proper variable to plot.  Positive values are contoured with full
lines, and negative values are contoured with dashed lines.  In
Figure~\ref{jupiter_side}, the maps are in orthographic projection
centered on $(\lambda,\phi) = (270,0)$ and gridded at 30-degree
intervals.  Note that with this projection the polar latitudes occupy
a smaller area of the disk than the low latitudes.  The resolution of
the simulation is T170---i.e., $N\! =\! 170$ in
Eq.~(\ref{basis})---and corresponds to a 512$\times$256
longitude-latitude grid over the globe.  This resolution is minimally
comparable to a 1024$\times$512 grid in a finite difference
calculation because of greater intrinsic accuracy of the
pseudospectral method. The duration of this run is 2000 planetary
rotations (i.e., $\tau\! =\! 2000$).  We have also performed shorter
duration runs at higher resolutions (T213 and T341) to verify
convergence of the results and find that there is no qualitative
difference in the evolution at the higher resolutions.  However, a
minimum of T63 (192$\times$96) resolution is required to resolve the
turbulence (i.e. growing vortices).
Time in unit of the planetary rotation periods is indicated in the
upper left corner of each frame.  The contour levels, 40 in all, are
identical in each frame.  In Figure~\ref{jupiter_top}, the maps are in
polar stereographic views centered on the north pole.  In this
projection also, the polar region occupies a smaller area of the disk
than the equatorial region.  East is in the counter-clockwise
direction, and $\lambda\! =\! 0$ is at the 3 o'clock position.

At $\tau\! =\! 0$, the atmosphere is stirred randomly and released
with a background corresponding to the planetary vorticity $f$.  By
$\tau\! =\! 84$, the formation of zonally elongated structures is
clearly evident at low latitudes.  These structures are formed by
growing eddies, which radiate Rossby (planetary) waves after reaching
the size of $L_{\beta}$ (see \S2).  Rossby waves are the undulations
of $q$ about a latitude circle.  Note that the wave propagation is
strongest at the equator since the linear dispersion relation from
Eq.~(\ref{PVC}) for the Rossby waves gives a longitudinal phase speed,
\begin{equation}
c_p\ =\ -\frac{\beta}{n(n+1)/R_p^2\, +\, 1/\mbox{\LR}_e^2}\, .
\end{equation}
Recall that $\beta\equiv R_p^{-1}df/d\phi = 2\Omega\cos\phi/R_p$ and
is a maximum at the equator.  The length, \LR$_e\equiv
(gH_e)^{1/2}/f$, is the ``equivalent depth'' Rossby deformation
radius, where $H_e$ is the equivalent depth---an important parameter
in atmospheric dynamics \citep{Gill82}, including in the theory of
tides \citep{Chapman70}.  In the equivalent-barotropic model, $H_e =
H_{p0}/{\cal A}_0$, a weighted pressure scale height evaluated at the
lower bounding surface.  Note also that \LR$_e$ is largest at the
equator and smallest at the pole, giving a small interaction length
for the vortices in the polar region (see $\tau\! =\! 423$ frame in
Figure~\ref{jupiter_top}, for example).  The small interaction length
significantly reduces motion and mergers in that region.  From our
simulations, we expect the polar region at the cloudtop level of
Jupiter to contain spots, rather than bands.  And, in fact, bands are
not observed above about 60 degrees latitude on Jupiter.

Maps of $q$ can be effectively used to study atmospheric circulation.
Such maps constitute a standard tool for studying the circulation on
the Earth \citep{Hoskins85}.  They have frequently provided insights
to important processes which are otherwise difficult to obtain.  Since
$q = q({\bf v})$, under certain ``balanced'' conditions (e.g.,
pressure gradient force balanced by the Coriolis force\footnote{This
condition is known as {\it geostrophic balance} \citep{Holton92} and
applies to Solar System giant planets, as well as the Earth, to a good
degree away from the equator.}), $q$ can be inverted to
obtain the wind field, ${\bf v} = {\bf v}(q)$.  That is, the wind
information is contained in $q$.  In spectral space, $q$ can be seen
to have a broad distribution.  This demands a high resolution for
accurate representation, both observationally and numerically.  Loss
of $q$ due to under-resolving, therefore, causes loss of pertinent
information.  Indeed, the ability to retain $q$ is one strong
justification for using the equivalent-barotropic model and also an
important advantage of our model compared with models using the finite
difference method.

Finely resolved $q$ features can be used to identify the presence of
atmospheric flow structures (e.g., vortices and jets), as well as
flows around the structures, to exquisite detail (cf., \citet{Cho01},
and references therein).  For identifying jets, consider a simple 2-D
zonal flow: $u = u(\phi)$ and $v = 0$ with $u, v \in \Re^2$.  If
incompressible (non-divergent) in 2-D, $\cal H$ is constant.  Then,
$Dq/Dt = 0$ with $q = (\zeta + f)$, and $\cal H$ plays no dynamical
role.  Using the definition of $\zeta$ ($\equiv{\bf
  k}\!\cdot\!\nabla\!\times\!{\bf v}$), we have $\zeta = -R_p^{-1}\del
u / \del\phi$.  Note that if $R_o\ll 1$ and $\zeta\sim U/L$, then
$\zeta\ll f$.  Therefore, for rapidly-rotating planets (such as
Jupiter), the Coriolis parameter $f$ dominates over the relative
vorticity $\zeta$, especially at high latitudes.  In this case, $q$ is
mostly positive (negative) in the northern (southern) hemisphere, as
can readily be seen in Figure~\ref{jupiter_side}.  Recall that $f =
2\Omega\sin\phi$ and hence changes sign when crossing the equator.
The gradient of $q$ in the meridional direction is then
\begin{equation}\label{q_grad_sphere}
\frac{\del q}{\del\mu} = \frac{\del f}{\del\mu} - 
R_P^{-1}\frac{\del^2 U}{\del\mu^2}\, ,
\end{equation}
where $U = u\cos\phi$ and $\mu = \sin\phi$.  On the plane tangent to
the planetary surface, this is just
\begin{equation}\label{q_grad_plane}
\frac{\del q}{\del y} = \beta - \frac{\del^2 u}{\del y^2}\, ,
\end{equation}
where $\beta = \del f/\del y$ and $y$ is the northward direction.
According to Eqs.~(\ref{q_grad_sphere}) and (\ref{q_grad_plane}), jets
may be identified by their curvatures.  For example, in
Eq.~(\ref{q_grad_plane}), an eastward jet corresponds to ${\del^2
  u}/{\del y^2} < 0$ while a westward jet corresponds to ${\del^2
  u}/{\del y^2} > 0$.  Consequently, for a given value of $\beta$
(which is always positive), a large (positive) value of $q$ gradient
indicates an eastward jet in the flow while a small (negative)
gradient of $q$ indicates a westward jet.  Thus, an atmosphere
containing alternating eastward and westward jets show a corresponding
alternating tightening and relaxing of $q$-contours in the meridional
direction.  The correspondence can be seen in
Figures~\ref{jupiter_top} and \ref{jupiter_jets}.  The absolute value
of $\del q/ \del y$ (or $\del q/ \del\phi$) is a direct measure of the
sharpness of the jet, eastward or westward.

Figure~\ref{jupiter_jets} depicts the quasi-steady jets from the
Jupiter simulation presented in Figures~\ref{jupiter_side}
and~\ref{jupiter_top} (Model S1).  It shows a zonally-averaged 
(eastward) wind profile as a function of latitude, $[u]$, where
\begin{equation}
[u(\phi)]=\frac{1}{2\pi}\int_0^{2\pi} u(\phi, \lambda)\ d
\lambda\, .
\end{equation}
The jets are the peaks in the wind profile.  The time of the
simulation is $\tau\! =\! 300$, well after the profile has emerged
from the initial stirring.  The profile is steady and nearly identical
even at $\tau\! =\! 2000$.  In the figure, there are about half dozen
zonal jets in each hemisphere with amplitude $\sim$25~m~s$^{-1}$.
There is also a very strong equatorial jet of amplitude $\sim
-125$~m~s$^{-1}$.  Qualitatively, the number, widths, and absolute
magnitudes of jets are all similar to those observed on Jupiter.  As
in the previous shallow-water studies of Jupiter \citep{Cho96b},
however, the sign of the equatorial jet is opposite to that observed
on Jupiter, as well as on Saturn (not shown). The retrograde
(westward) equatorial jet in the simulation is due to the strong
presence of Rossby waves at the equator, as discussed above.  Clearly,
an additional driving mechanism is also required in the
equivalent-barotropic equations model to capture the proper sign of
the equatorial jet, as in the shallow-water equations model.  This is
not unexpected given the formal similarity between the
equivalent-barotropic equations and the shallow-water equations.
Later, in \S5, we return to the issue of prograde (eastward)
equatorial jet---currently an unsolved problem in planetary
atmospheric dynamics.  For now, we simply note that the model does
capture the proper sign of equatorial jets on Uranus and Neptune, as
we shall show shortly.  Evidently, some mechanism not included in the
simplest version of the equivalent-barotropic equations is required
for Jupiter (cf., \citet{Heimpel05}) and Saturn but not for Uranus and
Neptune.  We also note that simulated bands are more wavy and contain
fewer spots than on the observed planets.  However, these simulated
features appear to match the observed ones better with increased
resolution in the simulations.

In Figure~\ref{sw_compare}, we present a direct comparison of the
equivalent-barotropic and shallow-water models.  We do this to
demonstrate the validity of our model.  When $\kappa\!  =\!  1$, the
adiabatic equivalent-barotropic equations formally reduce to the
shallow-water equations, with ${\cal A}_0$ set to unity.\footnote{We
  remind the reader that vortex columns cannot tilt in the
  shallow-water model.  So, a unit ${\cal A}_0$ is physically
  consistent.}  In this case, $H_{p0}$ becomes the layer thickness in
the shallow-water model (Eq.~(\ref{thickness})).  Note that, while
formally equivalent, the ``shallow-water limit'' of
equivalent-barotropic equations does not carry a physical
interpretation.  This is because physically $\kappa\!  =\! 1$
corresponds to $c_v\!  =\! 0$ and $\gamma \rightarrow \infty$, where
$c_v$ is the specific heat at constant volume and $\gamma = c_p/c_v$
(recall that $\kappa = R/C_p$).  In the figure, two runs are
presented: on the left (a) is a frame from the equivalent-barotropic
calculation and on the right (b) is a frame from the shallow-water
calculation (Models J1 and J2 in Table~\ref{tab:two}, respectively).
In the two runs, all parameters are identical, except that the
equivalent-barotropic equations model calculation on the left has
$\kappa\! =\! {\cal A}_0\! =\! 1$.  The resolution of the runs is
T106.  The projections of the maps are polar stereographic, as in
Figure~\ref{jupiter_top}.  The time of the frame is at $\tau\! =\!
179$, long after the quasi-steady state has been reached.  In these
two runs, the initial spectrum is peaked at a lower wavenumber (larger
scale) than in the run presented in Figures~\ref{jupiter_side} and
\ref{jupiter_top}.  As a consequence, the steady state is reached
earlier in these runs, compared to the run in
Figures~\ref{jupiter_side} and \ref{jupiter_top}.  However, because
both runs start with peaks at scales much smaller than the Rhines
scale $L_\beta$, there is no qualitative difference in the final flow
configuration: only the time when the equilibrium state is reached is
different.  Significantly, the results from the two models, (a) and
(b), are practically identical.\footnote{They of course cannot be
  exactly identical because of the initial random seeding and minor
  numerical differences.}  This verifies that flow features are truly
robust (independent of the randomness of stirring, for example) and
are reproducible by the new model.

Figure~\ref{neptune} shows three successive equatorial orthogonal (a)
and polar stereographic (b) views of a typical Neptune simulation from
our model (Model S4 in Table~\ref{tab:two}).  In the figure, three
time frames are shown to illustrate {\it i)} the flow in the initial
state ($\tau\! =\! 0$), {\it ii)} during transition to the
quasi-steady equilibrium state ($\tau\! =\! 33$), and {\it iii)} the
equilibrium state itself ($\tau\! =\! 146$).  As in
Figures~\ref{jupiter_side} and \ref{jupiter_top}, iso-$q$ contours are
plotted with positive (negatives) values in full (dashed) lines.  The
initial spectral distribution is the same as in that of
Figures~\ref{jupiter_side} and \ref{jupiter_top}, except $\bar U$ is
larger (see Table~\ref{tab:one}).  In this simulation, the structures
dominated by scales around the wavenumber $n\sim14$ at $\tau\! =\! 0$
are more easily recognized than in the Jupiter case because of the
larger $\bar{U}$ and smaller $f$ for Neptune.  More importantly,
because of the larger $\bar{U}$ and smaller $f$, the flow quickly
evolves into a final configuration dominated by two strong circumpolar
vortices (one centered on each pole) and a broad jet, associated with
a well-homogenized $q$ region at low latitudes.  In fact, only a few
$q$-contours are seen all the way up to the high latitudes, until the
boundaries of the polar vortices are reached.  Such a $q\! =\! 0$
low-latitude region was previously {\it assumed} in modeling the
chaotic, wobbling motion of Neptune's Great Dark Spot with great
success \citep{Polvani90}.  The strong polar vortex results from
vigorous mergers, in which surrounding vortices are continuously
``soaked up'' before they are sheared away or dissipated.
Dynamically, the strong interaction is expected given the large value
of \LR$/R_p\! \sim\! 1/6$ for Neptune, in contrast with $\sim$1/30 for
Jupiter (see Table~\ref{tab:two}).  Note that the circumpolar vortices
are cyclonic.  Under geostrophic balance, the cyclonic vortex is
associated with a cooler region than its surroundings.

In the case of Neptune as well, the steady-state configuration
(already well in place after $\tau\!\sim\! 30$) is stable over
timescales of several thousand $\tau$.  Significantly, as will be
shown shortly, Neptune's jet profile is closer to that we obtain for
close-in extrasolar giant planets than Jupiter, as reported in
\citet{Cho03}.  This marked difference mainly arises from the lower
$\Omega$ values of close-in extrasolar giant planets compared to that
of Jupiter.  In fact, assuming synchronization, close-in extrasolar
giant planets rotate only moderately fast compared to Solar System
giant planets.  Additional differences between close-in extrasolar
giant planets and Neptune also result from the intense irradiation
experienced by the close-in extrasolar giant planets, as we shall see
below (\S5).

But, first we quickly summarize our equivalent-barotropic equations
results for the Solar System giant planets.  Figure~\ref{ss_jets}
shows the zonal wind profiles at late times obtained in our
simulations of all four Solar System giant planets (Models S1--S4 in
Table~\ref{tab:two}): Jupiter~(a), Saturn~(b), Uranus~(c), and
Neptune~(d).  Qualitatively, these profiles match well the observed
profiles in terms of the overall number of bands and strengths of jets
on each planet (see Cho \& Polvani 1996b for observed profiles).
Generally, the four planets organize themselves into two discernible
groups in terms of their general circulation patterns at the cloudtop
levels.  Jupiter and Saturn share similar flow patterns, with their
multiple jets and narrow bands; and, Uranus and Neptune share similar
flow patterns, with their few jets and broad bands.

In the current study, we have performed over 300 simulations in order
to thoroughly explore the numerical and physical parameter-space
spanned by the equivalent-barotropic equations system.  Based on this
extensive exploration, it appears that capturing the prograde rotation
on Jupiter and Saturn requires physics beyond that included in the
current, adiabatic equivalent-barotropic equations model.  We give one
example of a possible extension in \S5.  It is crucial to note that
the lack of correspondence is not simply an issue of including the
third dimension.  One well-known mechanism in 3-D situation which does
produce prograde rotation in simulations of the Earth \citep{Suarez92}
is a zonally-asymmetric ($m\!  =\! 1$ and $m\! =\! 2$),
large-amplitude forcing that is symmetric in the meridional direction.
However, this mechanism does not apply to Jupiter (since it is not
thought to be zonally asymmetrically heated) and cannot explain Saturn
(since it is not meridionally symmetrically heated).  In
\S5, we discuss the consequences of this forcing further for close-in
extrasolar giant planets, where it may find some justification.

In summary, the results from our study illustrate several generic
features of giant planet circulations modeled with the
equivalent-barotropic equations.  First, there is no qualitative
difference between model results for equivalent-barotropic equations
and shallow-water equations.  Second, with the knowledge of a few
global physical parameters, we are in a position to capture the gross
properties of unsynchronized extrasolar giant planet atmospheric flows
(several examples of possible unsynchronized extrasolar giant planet
circulations are presented in the ensuing sections).  Third, as in the
previous shallow-water and equivalent-barotropic studies, the scales
$L_{\beta}$ and \LR\ emerge as critical dynamical parameters for
effecting realistic ``predictions'' of general circulations.  Last, a
knowledge of the strength and distribution of {\it diabatic} forcing
(i.e., not just incident but also baroclinic, convective, latent heat,
etc.) appears minimally needed to properly model the direction of the
equatorial jet.

\subsection{A Warm Jupiter}

Figure~\ref{warmer_jupiter} illustrates our simulation of a ``warm''
Jupiter (Model U1 in Table~\ref{tab:two}).  This may serve as an
example of a Jupiter-like extrasolar giant planet which has migrated
in some distance toward its host star but is not yet synchronized.  At
the same time, since the temperature increases with depth below the
main cloud deck on Jupiter, it may heuristically serve as a
representative flow at a deeper level in the Solar System planet.  In
this simulation, all the parameters are identical to that shown in
Figures~\ref{jupiter_side} and \ref{jupiter_top} (Model~S1), except
$\bar{\cal H}$ in this simulation is such that the global average
temperature $\bar{T}_0 = 300$~K (roughly a ``Jupiter at 1 AU'').  Both
stereographic (a) and orthographic~(b) views are shown at $\tau\! =\!
199$, well after the formation of steady bands and jets.  This figure
should be compared with the latter time frames of
Figures~\ref{jupiter_side} and \ref{jupiter_top} (Model S1).

In the early part of the evolution, the flow is very similar to the
early evolution of Model S1.  However, small differences begin to
appear as the evolution proceeds: the banding at low to mid latitudes
is weaker than that in Model~S1 and even shows signs of slight erosion
at times.  The effect of the erosion can be seen in
Figure~\ref{warmer_jupiter}, where the positive gradient of $q$ at low
latitudes is weaker compared with those in Figures~\ref{jupiter_side}
and \ref{jupiter_top}.  The result is that bands generally appear more
diffuse and there are one or two fewer jets in each hemisphere.  The
amplitude of the equatorial jet is reduced by more than 50\% as well.
This behavior is consistent with the findings in \citet{Cho96a} for
the shallow-water case and is due to the increase in $\bar{\cal H}$
leading to a larger \LR\ (Rossby deformation radius).  The larger \LR\
acts as increased ``stiffness'', or reduced horizontal
compressibility; this leads to stronger interaction between the
structures in the fluid and the bands themselves undergo several
``mergers'', reminiscent of the vortices.  Formally, the shallow-water
equations reduce to 2-D, incompressible equations in the limit of
\LR$\rightarrow\infty$.  Note that in the figure no mid-latitude
anticyclonic spots survive and only a few vortices are present in the
polar region, as a further consequence of the larger \LR.

Given the behavior illustrated in Figure~\ref{warmer_jupiter}, we may
draw several general conclusions.  First, a much closer-in Jupiter may
have a slightly different visual appearance than the present Jupiter
at 5~AU.  It will possess broad, diffuse bands with very few, possibly
no, spots at the main cloud deck.  Second, except for the equatorial
jet, the overall jet structure is not changed very much, attesting to
the strong control rotation has on the flow.  The ratio, \LR/$R_p$, is
still $\ll 1/3$ in this simulation and a complete breakdown of the
basic jet structure is not expected: an unrealistically large
$\bar{T}_0 \gsim 1.3\!\times\! 10^4$~K is necessary for the breakdown.
Finally, given that the situation can also be considered as crudely
modeling a deeper level (one pressure scale height lower) in the
current Jupiter, the flow is fairly vertically coherent (i.e.,
barotropic).  There is not much difference in the flow compared with
that at the ``higher level'' (i.e., flow with $\bar{T}_0 = 130$~K),
even in the absence of the explicit vertical coupling present in the
baroclinic models.

\subsection{Unsynchronized \HDblah}

Having established the singular importance of the rotation rate
$\Omega$, by considering Jupiters with two different equilibrium
temperatures (or orbital radii), we now consider the case of a
``Jupiter'' when its $\Omega$ is significantly reduced (e.g., by tidal
interaction) but still not locked in 1:1 spin-orbit resonance.  Such a
planet would have a small semi-major axis, $a\lsim 0.2$~AU.  This is
an important case to study since it allows a clear delineation of the
consequences of tidal locking on the atmospheric circulation.  It is
also relevant for the recently proposed scenario for \HDblah\ caught
in a Cassini state \citep{Winn05}.  As such, Figure~6 presents one of
the key results of this study: atmospheric circulation on a planet
with parameters representative of \HDblah\ (see Table~\ref{tab:one})
but without the explicit asymmetric forcing expected in the case of
tidal synchronization.

In Figures~\ref{unsync_hdblah_early} and \ref{unsync_hdblah_late}, the
flow evolution from Model~U2 (see Table~\ref{tab:two}) is shown at the
early stage (a) and at the equilibrated stage (b), respectively.  Once
the latter stage is reached, the flow state does not change in its
essential character: the circulation is dominated by a single,
coherent vortex at the pole and a sharp jet at its periphery,
reminiscent of those on Neptune/Uranus rather than Jupiter/Saturn
(compare with Figures~\ref{jupiter_top} and \ref{neptune}).  As in the
simulations of Solar System giant planets, the initialization here is
a random stirring, which is nonlinearly balanced and consistent with a
specified $T_0$ and $\bar{U}$.  In this simulation, $\bar{T}_0 =
1450$~K and $\bar{U} = 400$~m~s$^{-1}$, plausible values for \HDblah.
The actual values, particularly $\bar{U}$, are currently unknown.
Note that these values give \LR/$R_p\sim 1$ and $L_\beta/R_p\sim 1$,
as shown in Table~\ref{tab:two}.

Although clearly more similar to the circulation of Neptune/Uranus,
there are some distinguishable differences.  The polar vortex in
Figure~\ref{unsync_hdblah_late} is much larger than in the Neptune
case.  Also, the vortex is {\it not} centered on the pole, as it is on
Neptune (see $\tau\! =\! 146$ frame in Figure~\ref{neptune}).
Instead, it can be seen revolving around the pole on a timescale of
about 2.7 planetary rotation periods in this run.  In general, the
timescale of the revolution around the pole depends on the strength
and position of the polar jet (related to $\bar U$), since the polar
vortex essentially moves with the jet (background zonal flow).  In our
extensive survey, we have found that the timescale of the revolution
around the pole range from a few days to several tens of days.  In
addition, comparing the $\tau\!  =\! 146$ frame in
Figure~\ref{neptune} with any of the frames in
Figure~\ref{unsync_hdblah_late}, one can identify a marked reduction
of Rossby wave amplitudes (undulations of $q$ lines about a latitude
circle) at low-latitudes in the extrasolar planet case---although wave
breaking (lateral overturning of a large-amplitude wave) occurs on
both planets.  Associated with this feature is the reduced mixing and
homogenization at low-latitudes of this planet, compared with Neptune.
All of these behaviors can be understood in terms of the larger \LR\
and smaller $\beta$ on the unsynchronized close-in extrasolar giant
planet.  Larger \LR\ means more robust mergers and a more dynamic
final vortex.  Smaller $\beta$ entails a weaker shearing environment
for vortices at low latitudes and smaller wave amplitudes there.

An off-the-pole polar vortex, as in Figure~\ref{unsync_hdblah_late},
has several significant consequences for mixing of heat and vertical
motion at high latitudes.  If the vortex is cyclonic, as in the
figure, the air in its interior is cooler than it's surroundings.
Hence, a translating vortex carries with it a large mass of cold air
around the pole.  Moreover, as the vortex translates, it pulls up warm
air from the lower latitudes ahead of it while pulling down cold air
from higher latitudes behind it.  This induces a poleward flux of heat
and meridional mixing.  The effect is accentuated by its meandering
motion about a latitude circle.  At the same time, the vortex cools in
the net because it is generally warmer than the radiative equilibrium
temperature (recall that very little stellar irradiation reaches the
high latitudes).  Hence, by the first law of thermodynamics, the
potential temperature of the air inside the vortex decreases.  But,
since positive stability of a stratified region implies
$\del\theta/\del z > 0$, where $\theta(z)$ is the basic potential
temperature, the cooling air must sink across $\theta$-surfaces to a
lower valued surface; that is, there is a slow, down-welling motion
inside the vortex (not modeled in this work).  In addition, planetary
waves further increase the downward drift by a hysteretic mechanism in
diabatic conditions \citep{Salby96}.  Later, in \S5, we will show that
planetary waves are induced by thermal forcing on extrasolar planets.

Figure~\ref{unsync_hdblah_jets}a shows the quasi-steady jet profile
from the simulation in Figures~\ref{unsync_hdblah_early} and
\ref{unsync_hdblah_late} (Model~U2).  In this simulation, in which
$\bar{U} = 400$~m~s$^{-1}$ has been adopted, three broad jets form.
The low number of jets is also a robust feature of our \HDblah\
simulations with day-night forcing (\S5.1).  While the magnitude and
sign of individual jets vary from run to run, the number and root mean
square speed of the jets mainly depend on $\bar{U}$, in rough
agreement with $L_\beta$.  Occasionally, when $\bar{U} \gsim
400$~m~s$^{-1}$, a two-jet profile forms.  The adopted $\bar{U}$ value
is characteristic of Saturn, Uranus, and Neptune (see
Table~\ref{tab:one}).  On extrasolar giant planets, $\bar{U}$ can
plausibly vary from $\sim$100~m~s$^{-1}$ to $\sim$1000~m~s$^{-1}$.  A
value several times greater than $\sim$1000~m~s$^{-1}$ is not likely
to be realistic for large-scale flows on extrasolar giant planets
since the sound speed $c_s \lsim 3000$~m~s$^{-1}$ ($\approx\!
2700$~m~s$^{-1}$ on \HDblah).\footnote{This need not be so for
  small-scale waves, which may propagate with speeds close to or at
  $c_s$.}  The two-jet configuration also forms in the special case
when $\bar{U} = 0$, as we will show in \S5.3.  At the low end of the
range, when $\bar{U} = 100$~m~s$^{-1}$, a four or five jet profile can
result, consistent with $L_\beta$.  This case is illustrated in
Figure~\ref{unsync_hdblah_jets}b.

At this point we wish to remark on an issue pertaining to jets which
requires clarification.  There appears to be some confusion in the
current literature concerning the universality of retrograde
equatorial jets in one-layer turbulence calculations.  We emphasize
here that {\it a retrograde equatorial jet is not a necessary outcome
of all shallow-layer turbulence calculations}.  That outcome requires
a small \LR, as pointed out in \citet{Cho96a}.  In the case of Solar
System giant planets, \LR\ is indeed small for all the planets, and
the simulated equatorial jet is always retrograde in the absence of
forcing.  On extrasolar giant planets, \LR\ need not be small.  In
fact, \LR\ is generally not small for close-in extrasolar giant
planets because they are much hotter and less-rapidly rotating than
Solar System giant planets.  Accordingly, the equatorial jet in
Figure~\ref{unsync_hdblah_jets}b is actually prograde.  We have
performed many simulations of \HDblah\ under a variety of conditions
and find that what is robust is the adherence of the general flow
pattern to $L_\beta$, {\it not} the direction of any particular zonal
jet.

As already discussed, another dynamically relevant factor that is
currently unknown is the radiative equilibrium temperature
distribution on extrasolar giant planets.  The distribution is related
to the stellar irradiation but also depends in a complicated way on
the composition, cloud and aerosol distribution, and the dynamics
itself through the role it plays on the distribution of the
radiatively active components.  Figure~\ref{unsync_hdblah_temp} shows
the simulation presented in Figure~6 (Model~U2) at two different
equilibrium temperatures: (a)~$1800$~K and (b)~$800$~K (Models~U3 and
U4, respectively).  The range of the two temperatures is broad and is
meant to roughly bracket $\bar{T}_0$ resulting from a variety of
physical situations (e.g., distance, stellar flux, and albedo).  With
the same flux and albedo assumed in Model~U2, the higher $T_0$ can
also be applied to the case of an \HDblah-like planet which is
$\sim$0.005~AU closer to the host star than \HDblah\ at present ($a\!
=\! 0.045$~AU).  Similarly, the lower $T_0$ can be applied to an
\HDblah-like planet that is $\sim$0.15~AU further away.  As can be
seen in Figure~\ref{unsync_hdblah_temp}, the higher $T_0$ leads to a
more smoothly varying low- to mid-latitude $q$ distribution while the
lower $T_0$ leads to much more planetary (Rossby) wave activity, along
with a more centered polar vortex.  In both cases, the basic three-jet
profile is unchanged.  However, interestingly, the higher $T_0$ value
leads to displaced a polar vortex, which may potentially be of
observational relevance \citep{Cho03}.

\section{Synchronized Extrasolar Giant Planets}

\subsection{Effects of Zonally Asymmetric Heating}

From an atmospheric hydrodynamics point of view, \HDblah\ is
particularly interesting in that both its radius and mass (hence its
surface gravity) are accurately known and it has been the subject of
most observational studies to date.  The radius and mass information
is now available for more than a dozen extrasolar giant planets and
this number is expected to grow in the future.  As already noted, even
if synchronization is invoked to pin down $\Omega$ and a ``cloudless
atmosphere'' is considered to leave aside the complexities of
radiative issues (clouds, albedo, vertical distribution of species,
etc.), there is still no information available to constrain the
amplitude of thermal forcing $\eta$ and the global wind speed
$\bar{U}$.  These parameters crucially control the flow dynamics
through local changes of \LR\ and global changes of $L_\beta$.
Eddy/wave amplitudes and propagations are affected as well.
Experience with Solar System planets tells us that circulation and
radiative transfer mutually interact in such a way as to modify the
temperature structure away from the simple radiative and
radiative-convective equilibrium.  In this section, having considered
a ``\HDblah'' in an unsynchronized state, we now isolate the effects
of a specified value of $\eta$ on the flow.  In
\S\ref{ubar_variation}, we present our findings of the effects of
varying $\bar{U}$ on the flow and the horizontal temperature
structure.  We wish to focus on the robustness of plausible
circulation patterns formed under turbulent conditions.

As a first study of the forcing effect, we assume a simple day-night
equilibrium temperature distribution, $z^\prime_0 \propto
(\cos\lambda\cos\phi)^\varkappa$, where $\varkappa$ is chosen from the
set of \{1/4, 1, 2\} for different runs.  Note that a full 1-D
radiative transfer calculation of \HDblah\ atmosphere applied over the
planet suggests that this is a good range for the planet, with
$\varkappa\! =\!  1/4$ being more appropriate for the upper levels and
$\varkappa\! =\!  1$ being more appropriate for the lower levels in
the stable layer of a close-in extrasolar giant planet.  There is no
qualitative difference in our results due to a different choice of
$\varkappa$ value adopted, however.  The main result of our study of
the forcing effect is as follows: in adiabatic, equivalent-barotropic,
turbulent simulations, the global circulation is not qualitatively
affected by the applied thermal forcing.  That is, the general
characteristics of global circulation (i.e., number of bands, strength
of zonal winds, presence/absence of dominant polar vortices) are still
well characterized by \LR\ and $L_\beta$ (or equivalently $R_o$ and
$B_u$), irrespective of the value of $\eta$ adopted.  This may appear
surprising at first, but one of the key points of the present work is
that the basic flow pattern is set by the dynamical parameters,
$L_\beta$, \LR, and $R_p$.  Unless very strong diabatic forcing places
the extrasolar planet into a special dynamical regime (see \S5.3),
forcing which merely deflects the isentropic surfaces (as opposed to
causing large mass to cross the surfaces, as in overturning surfaces)
cannot in itself produce major changes in the {\it flow} at this level
of modeling.  It can, of course, affect the {\it temperature}
structure, as will be shown shortly.

The robustness of the basic flow is illustrated in
Figure~\ref{forcing_compare_flow}.  The figure presents four T106 runs
(Models~H1, H2, H3, and H4 in Table~\ref{tab:two}), which are
identical in all respects except for the forcing perturbation
amplitude, $\bar{z}_0$.  In the figure, $\bar{z}_0$ spans from 0 to
20\% of the background temperature (i.e., $\eta\! =\! 0$ to 0.2),
corresponding to 0 to $\sim$300~K.\footnote{Recall that the average
  background is represented by average thickness (mass) of the layer.}
The latter is not a small deflection.  We have also run cases in which
the fractional deflections are even larger ($\eta\! =\! 0.4$), but the
behavior is basically a more violent version of the 20\% case.  Frames
at the same time ($\tau\! =\! 95$) from simulations with $\eta$ values
of 0 (a), .02 (b), .04 (c), and .20 (d) are shown.  As seen from the
figure, the forcing primarily amplifies the mid-latitude jet.  As
$\eta$ increases, low latitude $q$ is homogenized, signifying
strengthening of the retrograde equatorial jet (a--c).  Concurrently,
the $q$-gradient at mid-latitude, hence the jet there, becomes
sharper.  The wave breaking activity in the mid-latitudes also
diminishes, until the forcing is no longer small (a--c).  In the
large-amplitude case~(d), the jet becomes strongly zonally asymmetric,
becoming unstable, along with increased wave breaking activity in the
mid-latitudes.  When $\del q/\del\mu$ changes sign, due to the
sharpening of the jet, the jet meets the necessary condition for
barotropic instability and the unstable flow ensues \citep{Charney62}.

Figure~\ref{forcing_compare_temp} shows the effects of increased
forcing amplitude $\eta$ on the temperature distribution, $T_0 =
T_0(\lambda,\phi)$, at dynamical equilibrium.  The same runs in
Figure~\ref{forcing_compare_flow} (Models~H1, H2, H3, and H4) are
shown with $\eta$: (a) 0., (b) .02, (c) .04, and (d) .20.  As can be
seen, for a fixed $\bar{U}$, $T_0$ is strongly dependent on $\eta$.
With no or low forcing, $T_0$ essentially tracks the flow, the polar
vortex~(a).  As the forcing amplitude increases, however, competition
between the flow-induced moving temperature pattern and the
forcing-induced fixed temperature pattern develops: temperature
anomalies on the whole either rotate quasi-steadily about the pole~(b)
or oscillate aperiodically about a point near the substellar
point~(c).  When the amplitude is high, the forcing pattern dominates
entirely and $T_0$ is essentially the imposed day-night distribution
fixed on the substellar point~(d).  We emphasize here that the global
circulation is not modified much, at least in the adiabatic case.
Similar jet profiles and polar flow structures are present, even under
strong forcing (compare Figure~\ref{forcing_compare_temp} with
Figure~\ref{forcing_compare_flow}): an asymmetric\footnote{The cyclone
  is much stronger than the anticyclone.} cyclone/anticyclone pair at
high latitudes is always present (see
Figures~\ref{forcing_compare_flow} and \ref{cyclone_chop}), but it may
not be discernible in temperature maps if swamped by the applied
thermal forcing.  We note that when the same set of runs is repeated
with larger $\bar{U}$, the same trends in temperature maps are
exhibited at comparatively larger $\eta$ values, as expected from
stronger motion-induced temperature patterns.  Hence, thermal
variability depends on $\bar U$, as well as $\eta$, as will be further
demonstrated in the next subsection.  This is one of our key results.

In summary, clearly, the entire spectrum of temperature distribution
behavior is possible---from forcing being completely overwhelmed by
the flow to forcing completely overwhelming the flow.  Not
unexpectedly, the critical $\eta$ value at which $T_0$ makes a
transition to merely expressing the specified forcing depends on the
given $\bar{U}$ value.  The fact that the flow is fairly similar in
all these cases indicates that the coupling between the flow and
applied heating is relatively weak.  Diabatic forcing appears to be
required for a stronger coupling.  The overall implication here is
that, one cannot infer $\bar{U}$ and $\eta$ independently of each
other from a temperature map, at least in adiabatic conditions.

\subsection{Global Mean Kinetic Energy Variation}
\label{ubar_variation}

Figure~\ref{ubar_compare_flow} shows the dependence of the global
circulation on $\bar{U}$.  The figure presents four T106 runs
(Models~H1, H5, H6, and H7 in Table~\ref{tab:two}) that are identical
in all respects, except for $\bar{U}$.  In these runs, $\bar{U}$ spans
from 100 to 1000~m~s$^{-1}$ from run to run.  As noted already, the
latter is not a small value: globally-speaking, the circulation is no
longer in the quasi-geostrophic regime, when $\bar{U}\!\sim\!
1000$~m~s$^{-1}$, since $R_o\!\sim\! 1$.  Locally, of course, $R_o$
may be $\ll 1$, since the velocity varies over the planet.  In the
figure, simulations with $\bar{U}$ values of 100 m s$^{-1}$ (a), 200 m
s$^{-1}$ (b), 400 m s$^{-1}$~(c), and 1000~m~s$^{-1}$ (d) are
presented at the same time frame ($\tau\! =\! 95$).  We have also run
cases in which $\bar{U}$ is even larger (2000~m~s$^{-1}$), but in this
case the behavior is similar to that of (d), at early times.  At later
times, the flow field blows up, due to the large amount of surface
gravity waves that naturally arise.\footnote{The blowing up can be
  prevented, to a certain extent, if an unrealistically large amount
  of artificial dissipation is applied or, equivalently, if the
  resolution is very low.}  It can be seen clearly in
Figure~\ref{ubar_compare_flow} that increasing $\bar{U}$ increases the
size and strength of the polar vortex.  In fact, the vortex is large
enough in (d) to nearly cap the planet poleward of 60~degrees if it
were centered at the pole.  This results from the greater root mean
square kinetic energy, $\onehalf\bar{U}^2$, of the initial eddies.

It may appear to the reader that, in Figure~\ref{ubar_compare_flow},
planetary (Rossby) wave breaking activity abruptly terminates at high
$\bar{U}$.  However, in actuality, it increases.  This is demonstrated
in Figure~\ref{cyclone_chop}a--c, in which the field presented in
Figure~\ref{ubar_compare_flow}d (reproduced in
Figure~\ref{cyclone_chop}a) is successively contoured at smaller range
and intervals (Figure~\ref{cyclone_chop}b--c); the number of contours
is the same.  The vortex is ``chopped off'' to reveal the surrounding
flow.  Clearly there is a large amount of structure outside the
dominant vortex.  In fact, the large amplitude of the planetary
(Rossby) waves and their breaking has produced an anticyclone
(Figure~\ref{cyclone_chop}b).  This anticyclonic vortex corresponds to
a warm region compared to the interstitial region between the pair,
giving rise to an associated slowly-rotating thermal dipole.  This is
remarkably similar to what happens on the Earth: in the Earth's
stratosphere, upwardly-propagating planetary waves cause the
stratospheric polar vortex to be displaced from the pole and induce
the formation of an anticyclone, conserving $q$.  The interstitial
region itself is well mixed due to pronounced wave breaking, as on the
Earth (cf., \citet{Andrews87}).

Figure~\ref{cyclone_chop}d--e shows the effect of increasing $\bar{U}$
on the zonal winds.  The zonal wind profile is plotted for two
simulations presented in Figure~\ref{ubar_compare_flow}: (a) and
(d)---Models~H1 and H7, respectively.  The figure clearly shows that,
for fixed $\Omega$ and $R_p$, the number and width of the jets depend
on $\bar{U}$ in good agreement with the $L_\beta$ estimate.  Compare
also this figure with Figure~\ref{unsync_hdblah_jets}, in which
asymmetric heating was not applied.  From the comparison, one may draw
two general conclusions: 1) forcing does not appear to have a
distinguishable effect on the general structure of the zonal jets,
their number and width; and 2) given the broad range of $\bar{U}$
values in the simulations of the two figures, the low number of zonal
jets is a robust feature on \HDblah-like planets.

The $\bar{U}$ values can be shown to be intimately tied with the
resulting temperature distribution, as shown in
Figure~\ref{ubar_compare_temp}.  The same runs as in
Figure~\ref{ubar_compare_flow} (Models~H1, H5, H6, and H7) are shown
with $\bar{U}$ values: (a)~100~m~s$^{-1}$, (b) 200~m~s$^{-1}$, (c)
400~m~s$^{-1}$, and (d) 1000~m~s$^{-1}$.  As can be seen, for a fixed
forcing amplitude $\eta$, $T_0$ is strongly dependent on $\bar{U}$.
This is very similar to the behavior illustrated in
Figure~\ref{forcing_compare_temp}.  With small $\bar{U}$, $T_0$
essentially expresses the day-night difference (a); that is, the flow
is not energetic enough to advect temperature in significant amounts.
As $\bar{U}$ increases, however, competition between the flow-induced
temperature pattern and the forcing-induced temperature pattern
develops: temperature anomalies on the whole either oscillate
aperiodically about a point near the substellar point (b) or rotate
quasi-steadily about the pole~(c).  When $\bar{U}$ is high, the flow
dominates and $T_0$ essentially reflects the distribution dictated by
the dominant vortices~(d).  Again, in each case, there is an
asymmetric cyclone/anticyclone pair at high latitudes, independent of
$\bar{U}$.  When the same set of runs is repeated with larger $\eta$
values, the same trend is exhibited, with the transition to flow
dominance occurring at larger $\bar{U}$ value.  The only time a
qualitative departure is obtained is in the special case when the
simulation is started at a resting state and driven wholly by the
applied day-night forcing (without any stirrings), the situation to
which we now turn.

\subsubsection{Quiescent Initial State: A Special Case}

At the current level of modeling, be it with the equivalent-barotropic
equations or the full primitive equations, one of the ingredients
necessary for accurate modeling and good understanding of extrasolar
giant planet atmospheric circulation is the physics of baroclinic
processes.  To be more precise, what is important is the inclusion of
the {\it effects} of the processes, not so much the actual processes
themselves.
This is the idea of parameterization.  In our usual model setup, the
atmosphere is stirred in the beginning.  The stirring is designed to
crudely represent a generic form of baroclinic or barotropic
processes: eddies and waves generated by small-scale convection or
baroclinic instability, for example.  Both of these generating
mechanisms are expected to exist on extrasolar giant planets.  The
thermal forcing should be another source of strong eddies and waves,
especially on close-in extrasolar giant planets.  Once created, the
eddies and waves transport and/or propagate crucial dynamical
quantities, such as heat and wave-activity density\footnote{For
  homogeneous fluids, wave-activity density is
  $[\onehalf\zeta^\prime]/[\del q/\del\mu]$, where $\zeta^\prime$ is
  the vorticity-eddy.  It is sometimes also called (minus) {\it
    pseudomomentum} in the literature.}  \citep{Andrews87}.  For
example, the action of waves greatly affect both the source and the
target region (where the waves are dissipated) via interaction with
the background flow (e.g., jets) present in the target region.  This
is one reason why we have emphasized resolution and laid particular
stress on the jet profiles, and flow structures in general, in this
study.

Recently, \citet{Showman02} and \citet{Cooper05} have performed
baroclinic simulations of close-in extrasolar giant planet atmospheres
using the primitive equations.  In these simulations, the atmosphere
initially at rest is driven by day-night forcing applied in the energy
equation.  As in some simulations of ``stripped-down'' Earth, which
employ the primitive equations and similar forcing (i.e., one that is
low-order and meridionally-symmetric, but zonally-asymmetric), they
obtain close-in extrasolar giant planet atmospheres which are
superrotating.  That is, the zonal wind at the equator is prograde:
$[u\, (\phi\!  =\! 0)] > 0$. In Earth simulations, the emergence of
superrotation depends nonlinearly on the amplitude of the forcing,
sometimes making an abrupt transition from subrotating to
superrotating state at some critical amplitude \citep{Suarez92}.  At
criticality, a bifurcation is observed, the flow exhibiting sometimes
superrotation and sometimes not, for an identical set of parameters.
In some cases, the superrotation appears to be in a stable equilibrium
state, as the atmosphere remains superrotating even when the forcing
is removed.  Exactly how this peculiar behavior arises is currently
not well-understood.  Feedbacks between eddy momentum, angular
momentum, Hadley cell, bottom friction, and vertical resolution appear
to be at play.  Certainly, vertical angular momentum exchanges between
atmospheric layers and horizontal exchanges within a layer is required
for superrotation.

In simulations of the Earth, superrotation has also been achieved, and
investigated, through direct angular momentum forcing
\citep{Saravanan93}.  As with many phenomena admitted by the full
primitive equations, superrotation of planetary atmospheres can also
be studied using the equivalent-barotropic equations in a similar way.
In this work (see also \citet{Shell04}), an equatorial torque is
applied to represent what is expected to happen to and within an
individual layer in a (multi-layer) baroclinic atmosphere.  This is
done by adding a forcing term,
\begin{equation}
{\cal F} = \tilde{\cal F}\cos^n\phi\, ,
\end{equation}
to the zonal momentum part of Eq.~(\ref{equivalent-barotropic
  equations}a).  Here, $n$ is an integer and is set to unity in the
runs presented in this paper; we have checked that narrower forcing,
given by $n =$ 2 or 4, leads to qualitatively similar results.  Note that
the forcing is zonally-symmetric, even though the underlying
baroclinic process modeled is zonally-asymmetric.  As already
discussed, the exact (and unknown) form of forcing need not be
specified and is actually not sought: our aim here is an assessment of
the evolution of an established superrotating equatorial jet and the
nature of its interaction with eddies in an idealized setting.
Accordingly, the forcing is assumed not to respond to the modeled
layer itself (i.e., no back-reaction); we have in mind an ``active''
modeled layer that overlies a massive ``abyssal'' interior, which
merely acts as a source of low angular momentum and friction (e.g.,
due to convective turbulence).  One can think of this forcing as
arising from some general zonally-asymmetric baroclinic process that
produce eddy-momentum convergence at low latitude.  A small amount of
linear drag, of the form $-\alpha u$, is added to balance the forcing.

The resulting flow from Model~Z1 in Table~\ref{tab:two} is depicted in
Figure~\ref{rest_forced}.  The resolution of the simulation is T106.
The iso-$q$ contours in cylindrical projection, centered on the
substellar point.  Three time frames are shown: $\tau = \{35, 40,
100\}$ (Figure~\ref{rest_forced}a--c, respectively).  The
corresponding zonal jet profiles in each time frames are shown as
well.  In this simulation, the force-dissipation balance is set so
that the maximum wind is $\sim$1000~m~s$^{-1}$, similar to the
simulation reported in \citet{Showman02} at the 5.6--8.1~bar pressure
level.  A quiescent initial state is adopted, as in \citet{Showman02}
and \citet{Cooper05}.  In the figure, the first thing to note is that
the flow is meridionally (north-south) symmetric, as in
\citet{Showman02}.  This is expected given the setup and indicates a
complete dominance of the applied forcing.  Particularly clearly
visible in Figure~\ref{rest_forced}a are the large Rossby waves that
have been excited by the forcing.  One can also clearly see the
superrotation ($[u\, (\phi\! =\! 0)] > 0$) that is produced by the
forcing in the corresponding jet profile.  However, as the Rossby
waves propagate meridionally, they interact with the background flow
to drag the mid-latitude mean flow back (in the westward direction),
simultaneously causing the equatorial mean flow (the source location)
to speed up.

The behavior follows a well known acceleration/deceleration mechanism
in atmospheric dynamics.  The dissipation occurs in the region where
the phase speed of the waves match the mean flow speed.  Since Rossby
waves propagate westward, the flow must be eastward for the mechanism
to operate.  Observe in Figure~\ref{rest_forced}b that, where the wave
has dissipated, large-scale eddies have formed.  The flow is at this
point very similar to that of \citet{Showman02} (see their Figure~6).
However, ultimately, the flow is not stable---at least in the angular
momentum forced equivalent-barotropic system simulated here
(Figure~\ref{rest_forced}c).  The resulting flow is very complex
\citep[compared to][]{Showman02}, but the hemispheric symmetry still
persists; the increased complexity is due to the higher resolution.
One can see the strong influence of the propagating waves in
Figure~\ref{rest_forced}c, after which the jet profile does not change
very much.  Except for the equatorial jet, the mean flow is wholly
westward from the time shortly before $\tau\! =\! 100$.  Clearly, the
strength of superrotation depends on the pre-existing background flow,
given the forcing amplitude.

Using the equivalent-barotropic model, we have also explored the
behavior of superrotating flow states under various initial
conditions.  Our equilibrated-state findings are summarized in
Figure~\ref{rest_compare}, which present results from Models~Z2, Z1,
and Z3 in Table~\ref{tab:two}.  The figure illustrates the importance
of allowing eddies to be present in the flow field.  The projections
for $q$-field maps on the left column are as in
Figure~\ref{rest_forced}.

Figure~\ref{rest_compare}a illustrates thermally forced ($\eta\! =\!
0.4$) flow without applied angular momentum forcing ($\tilde{\cal F}\!
=\! 0$).  The jet profile consists of a mid-latitude westward jet in
each hemisphere, consistent with spreading of $q$ lines at
mid-latitudes by the applied day-night forcing.  The flow is not very
``dynamic'' in this case and does not change much over time.  In
contrast, with both angular momentum and day-night forcings applied,
the flow, shown in Figure~\ref{rest_compare}b, becomes more complex
and dynamic, with the background mean flow interacting strongly with
radiating planetary (Rossby) waves.  Finally, we have in
Figure~\ref{rest_compare}c all three forcings (day-night, angular
momentum, and initial random stirring) included in the simulation.
Here, the flow is similar to the turbulent flow cases we have
presented throughout this paper, with $\bar{U}\!\ne\! 0$.  In this
case, there are three broad jets.  Because the $\bar{U}$ associated
with the stirring is 400~m~s$^{-1}$, which is larger than the maximum
jet amplitude without the stirring (see Figure~\ref{rest_compare}b),
the equatorial jet here is weakly retrograde.  Hence, it appears that
the strength of superrotation is sensitive to the vigor of eddies
present.  More broadly, the absence of vigorous eddies may thus be a
general limitation in dynamical simulations.

\section{Conclusion}

In this work, we have extensively explored global, turbulent,
adiabatic atmospheric dynamics of extrasolar giant planets in circular
orbits.  Extrasolar giant planets with eccentricity $ < 0.05$ comprise
a significant fraction of the currently known population.  Several
robust properties emerge from our set of simulations, which broadly
confirm the results in \citet{Cho03}.  First, there is a strong
rotational control on the flow dynamics.  This leads to atmospheric
flows with a strong zonal component on extrasolar giant planets,
independent of the thermal forcing amplitude.  Hence, the velocity
field is {\it not} one of a simple diverging flow away from the
substellar point.  This is so even for close-in extrasolar giant
planets.  This behavior is reminiscent of Uranus, which possesses
strong zonal winds like the other Solar System giant planets even
though it is heated at the pole.  Second, in trying to understand the
general circulation of extrasolar giant planet atmospheres, it is
clearly useful to divide the task along the following two lines: {\it
  i)} understand the zonally-symmetric and zonally asymmetric
circulation; and, {\it ii)} understand the tropical-subtropical and
midlatitude-polar dynamics.  Of course, each includes the interaction
with the other two components.  Third, in the adiabatic
equivalent-barotropic equations model, it is possible to have a full
range of global temperature distributions.  They include rotating,
oscillating, shifted, and fixed day-night distributions.

Not surprisingly, the spatiotemporal behavior exhibited by the
temperature distribution depends on the amplitude of the thermal
forcing.
When thermal forcing is strong, the day-night temperature difference
is expected to be fixed.  In effect, the temperature field decouples
from the dynamics and can essentially be thought of as ``thermal
orography''.  However, the degree of decoupling depends on the
strength of the background flow so that, if pre-existing zonal winds
and eddies are strong, the temperature field is again enslaved to the
flow field.  Therefore, it appears from the adiabatic calculations
performed in this work that it would be difficult to ascertain the
characteristic flow speed independently of the radiative properties
(e.g., albedo)---and vice versa.  Note that because not as much
irradiation is absorbed at higher altitudes (in the absence of special
absorbers), the smaller amplitude forcing situations can be considered
very roughly applicable to those regions.  This would lead to a
vertical shear in the temperature field and a baroclinic adjustment
might be indicated.  It would be interesting to study such adjustment
processes when vertical communication between different layers is
taken into account.

In general, the response of the atmospheric motion to both mechanical
and radiative types of forcing is very complex and difficult to
analyze from first principles.  Modeling such complexity requires a
range of approaches---from simple analytical calculations to full
general circulation modeling.  Different approaches are more
successful with different pieces of the full problem, and they are all
needed to make good progress (e.g.,
\citet{Showman02,Cho03,Menou03,Burkert05,Cooper05}).  In this work,
the equivalent-barotropic equations in isentropic coordinate have been
used to focus on lateral dynamics near the top of the planet's
convection zone.  The isentropic layers above this region do not lie
at constant geometric heights.  The layers slant slightly downward
from the substellar point to the poleward direction on extrasolar
giant planets; on close-in extrasolar giant planets the layers slant
downward toward the antistellar direction as well.  On the other hand,
the isentropic layers below the region (the top of the convection
zone) slant upward by a very significant amount from the substellar
point.  In general, air parcels move both along and across these
surfaces, transporting heat in both horizontal and vertical
directions.  We have taken advantage of the clear association of
horizontal and vertical motions with adiabatic and diabatic heating,
respectively, to perform equivalent-barotropic equations simulations
in what amounts to ``1\onehalf-layers'' (i.e., a layer with variable
thickness).

Despite the vertical integration, the equivalent-barotropic equations
support many of the phenomena and types of fluid motion supported by
the primitive equations.  This includes Rossby waves, gravity waves,
balanced motions (e.g. geostrophic), adjustments, and barotropic
instability.  Some of the consequences of baroclinic processes, such
as stirring by eddies or convection can also be represented and
spatially reasonably resolved.  The high resolution also allows mixing
of $q$ and fine-scale tracers and turbulent cascades to small
scales---all critical features in real atmospheres.  Consequences of
thermal forcing on eddies and mean-flow can be included through the
deflection of lower boundary height.  In addition, the
equivalent-barotropic equations have a consistent set of conservation
laws for mass, energy, angular momentum, $q$, potential enstrophy
(${\cal H}q^2$), and more exotic quantities like pseudomomentum (wave
activity).  In view of all these properties, we have carefully
explored the dynamical behavior of the adiabatic equivalent-barotropic
equations model in this work.

In subsequent studies, we will consider more general physical
situations important for understanding atmospheric dynamics on
extrasolar giant planets.  We will explore the consequences of
relaxing the important adiabatic and barotropic assumptions made in
the present work, by using models that explicitly include radiative
forcing and vertical coupling between multiple atmospheric layers.
With the recent breakthrough detections of infrared emission from
several close-in extrasolar giant planets, observational programs
should be able to provide better constraints in the future on key
atmospheric parameters, allowing more accurate assessments of the
circulation and thermal structures on these planets.

\acknowledgements This work was supported in part by NASA contracts
NAG5--13478 and NNG06GF55G.  JYKC thanks Ursula Wellen for helpful
comments on the manuscript.  We also thank the referees for comments
that led us to clarify the main assumptions made in this work.


\begin{table}
{\tiny
\caption{Global Planetary Parameters}
\begin{center}
\begin{tabular}{lcccccccccc} \hline \hline
\\
Planet$^{(1)}$ & $M_\star^{(2)}$ & $P_{\rm orb}^{(3)}$ & $a^{(4)}$ 
& $e^{(5)}$ & $M_p^{(6)}$ & $R_p^{(7)}$ & $g^{(8)}$ 
& $\Omega^{(9)}$ & $\bar{H}_p^{(10)}$ & $\bar{U}^{(11)}$ \\
 & ($M_\odot$) & (days) & (AU) & & ($M_{\rm J}$) & (m) & (m~s$^{-2}$) 
& (rad s~$^{-1}$) & (m) & (m~s$^{-1}$)\\
\\
\hline
\\
Jupiter & 1.0 & 4,332.6& 5.2 & 0.0489 & 1.0 & $7.1 \times 10^7$ & 23 & $1.8 \times 10^{-4}$ & $2 \times 10^4$ & 70 \\    
Saturn & 1.0 & 10,759.2& 9.58 &0.0565& 0.3 & $6 \times 10^7$ & 9 & $1.6 \times 10^{-4}$ & $4 \times 10^4$ & 400 \\
Uranus & 1.0 & 30,685.4& 19.2 &0.0457& 0.046 & $2.6 \times 10^7$ & 9 & $(-)1 \times 10^{-4}$ & $3.5 \times 10^4$ & 300 \\
Neptune & 1.0 & 60,189.0& 30.05 & 0.0113& 0.054 & $2.5 \times 10^7$ & 11 & $9.75 \times 10^{-5}$ & $3 \times 10^4$ & 300 \\
\\
\hline
\\
\HDblah  & 1.05 & 3.5247 & 0.045 & 0.0 & 0.69 & $10^8$ & 8 
& $2.1 \times 10^{-5}\, ^{(a)}$ & $7 \times 10^5\, ^{(b)}$ &--\\ 
\\
\hline
\end{tabular}
\label{tab:one}
\end{center}} \tablecomments{(1) Giant planets (2) Parent star mass (3)
Orbital period (4) Semi-major axis (5) Eccentricity (6) Planet mass
(7) Planet radius (8) Surface gravity (9) Rotation rate (10) Global
pressure scale height (11) Global root mean square velocity scale (a)
Assuming spin-orbit synchronization (b) From global radiative
equilibrium.}
\end{table}       

\clearpage

\begin{table}
{\small
\caption{Summary of Simulations Discussed}
\begin{center}
\begin{tabular}{cccccccl} \hline \hline
\\
Model & \hspace*{.2cm} $\bar{U}$$^{(1)}$ & \hspace*{.4cm} $\bar{T}_0$$^{(2)}$ 
      & $\eta\, ^{(3)}$ & \hspace*{.2cm} \LR$^{(4)}$ 
      & \hspace{.2cm} $L_\beta$$^{(5)}$ & N$_{\rm bands}^{(6,\dagger)}$ 
      & \hspace*{1.5cm} Remarks \\
      & (m s$^{-1}$)    &  (K) & & ($R_p$) & ($R_p$) & \\
\\
\hline
\\
S1 & 70   & 130   & 0    & .03  & .23 & 14 & Jupiter \\
S2 & 400  & 95    & 0    & .03  & .64 &  5 & Saturn ($q$-map not shown)\\
S3 & 300  & 60    & 0    & .11  & 1.1 &  3 & Uranus ($q$-map not shown)\\
S4 & 300  & 60    & 0    & .11  & 1.1 &  3 & Neptune \\
\\
J1 & 70   & 130   & 0    & .02  & .23 & 14 
   & S1 w/ different size initial stirring \\
J2 & 70   & \hspace*{.5mm} 130$^\ast$ & 0 & .02 
   & .23 & 14 & J1 w/ shallow-water model \\
\\
U1 & 70   & 260   & 0    & .05  & .23 & 14 & ``Warm'' Jupiter   \\
U2 & 400  & 1450  & 0    & 1.2  & 1.2 & 3  & ``Unsynchronized'' \HDblah \\
U3 & 400  & 1800  & 0    & 1.3  & 1.2 & 3  & ``Warm'' U2 \\
U4 & 400  & 800   & 0    & 0.9  & 1.2 & 3  & ``Cool'' U2 \\
\\
H1 & 100  & 1450  & .0   & 1.2  & 0.6 & 5  & \HDblah, no thermal contrast \\
H2 & 100  & 1450  & .02  & 1.2  & 0.6 & 5  & \HDblah, low thermal contrast \\
H3 & 100  & 1450  & .04  & 1.2  & 0.6 & 5  & \HDblah, medium contrast \\
H4 & 100  & 1450  & .20  & 1.2  & 0.6 & 5  & \HDblah, high thermal contrast \\
\\
H5 & 200  & 1450  & .04  & 1.2  & 0.8 & 4  & \HDblah, low speed jets \\
H6 & 400  & 1450  & .04  & 1.2  & 1.2 & 3  & \HDblah, medium speed jets \\
H7 & 1000 & 1450  & .04  & 1.2  & 1.8 & 2  & \HDblah, high speed jets \\
\\
Z1 & 0    & 1450  & .04  & 1.2  & 1.8$^\ddagger$  & 2 
   & H6 w/ zero initial vel. \& superrot. \\
Z2 & 0    & 1450  & .04  & 1.2  & 1.8$^\ddagger$  & 2 
   & Z2 w/o superrot. \\
Z3 & 0    & 1450  & .04  & 1.2  & 1.8$^\ddagger$  & 2 
   & Z1 w/ eddies \\
\\
\hline
\end{tabular}
\label{tab:two}
\end{center}} \tablecomments{(1) Initial global root mean square velocity (2)
Global mean temperature at cloudtop for Solar System giant planets and
fiducial cloudtop for extrasolar giant planets (3)~Forcing amplitude
(4)~Nondimensional Rossby deformation radius at the pole, based on
$\bar{H}_p$ (5) Nondimensional Rhines length at the equator, based on
$\bar{U}$ (6) Number of zonal jets estimated, based on $L_\beta$
($\dagger$) May eventually erode to 2, if \LR $\gsim 0.3$ ($\ast$)
Based on layer thickness ($\ddagger$) Based on $\bar{U}$ at
quasi-equilibrium.}
\end{table}

\clearpage

\begin{figure}
  \figurenum{1} 
  \caption{Equivalent-barotropic turbulence simulation of Jupiter.
    Contour maps of the flow tracer, potential vorticity $q$, from six
    time frames are shown in orthographic view centered at the
    equator.  Positive (negative) values are in full (dashed)
    contours; 40 contour levels are shown.  Time, in unit of planetary
    rotation periods ($\tau = 2\pi/\Omega$), is indicated in the upper
    left corner of each frame.  Using only five physical parameters,
    which are known from observations, the initially random turbulent
    flow self-organizes into one dominated by zonal (east-west)
    bands---alternating, high/low $q$ gradients in the meridional
    (north-south) direction---similar to the actual Jupiter.}
  \label{jupiter_side}
\end{figure}  

\begin{figure}
\figurenum{2} 
\caption{Simulation of Figure~\ref{jupiter_side} in stereographic
  view.  The boundary of the disk is the equator, and the center of
  the disk is the north pole.  In this projection, the polar region
  occupies a smaller area of the disk compared with the equatorial
  region, as in Figure~\ref{jupiter_side}.  A quasi-steady banded
  state is reached and is robust.  The general flow picture does not
  change even at time, $\tau\! =\! 2000$, the duration of this run.
  The interaction of the flow structures (vortices and jets) is weak,
  due to the extremely small Rossby deformation radius, \LR\ (see text
  for definition).  Jets and vortices do not coalesce.}
\label{jupiter_top}
\end{figure}  

\begin{figure}
\epsscale{.8}
\figurenum{3} 
\caption{Quasi-steady zonal jet (positive eastward) profile from the
  simulation of Figures~\ref{jupiter_side} and \ref{jupiter_top} at
  $\tau\! =\! 300$.  The jets are associated with the bands in
  Figures~\ref{jupiter_side} and \ref{jupiter_top}.  {Several
  qualitative} features (number, width, and strengths of jets) of the
  observed profile on Jupiter are reproduced, giving confidence in the
  model approach.  As in the shallow-water model calculations of
  \citet{Cho96b}, however, the sign of the equatorial jet is opposite
  to that observed.  This is a common feature of one-layer turbulent
  models of Solar System giant planets without forcing.  The observed
  prograde (eastward) equatorial jet is due to mechanism(s) not
  included in these unforced models.}
\label{jupiter_jets}
\epsscale{1.}
\end{figure}

\begin{figure}
\figurenum{4}
\caption{Detailed comparison with the shallow-water model.
  Shallow-water simulation (a) and corresponding equivalent-barotropic
  simulation (b) are essentially identical.  However, the
  equivalent-barotropic model is superior to the shallow-water model,
  since crucial parameters (e.g., the equivalent depth $H_e$) take on
  a more physical meaning.  Note the presence of a large, stable
  anticyclone (a vortex, which is counter-rotating with respect to
  $\Omega$, which is anti-clockwise in the figure) at mid latitude [at
  $\sim$315$^\circ$ in (a) and at $\sim$230$^\circ$ longitude in (b)
  at $\sim$40$^\circ$ latitude in both].  This structure is similar to
  Jupiter's Great Red Spot and is of approximately the same size and
  shape.  The structure, along with the bands and jets, also appears
  in Figure~\ref{jupiter_top}, which has a different initial
  condition.  Hence, these are robust features in our simulations.}
\label{sw_compare}
\end{figure}  

\clearpage

\thispagestyle{empty}
\setlength{\voffset}{-20mm}
\begin{figure}
\figurenum{5}
\epsscale{.8}
\caption{Simulation of Neptune.  In contrast to Jupiter (e.g.,
  Figures~\ref{jupiter_side} and \ref{jupiter_top}), there is a large
  band of homogenized potential vorticity around the equator and a
  strong circumpolar vortex centered at each pole ($\tau\! =\!  146$
  frame).  A very small number of contours at low latitudes is
  present, compared with the high latitudes.  The overall flow pattern
  is a direct consequence of the larger $\bar{U}$ and smaller $\Omega$
  values (i.e., larger $\mbox{\LR}/R_p$ and $L_\beta/R_p$,
  respectively), compared to those of Jupiter. }
\label{neptune}
\epsscale{1.}
\end{figure}  

\clearpage

\setlength{\voffset}{0mm}

\begin{figure}
\figurenum{6}
\caption{Zonal jet profiles of the Solar System giant planets: (a)
  Jupiter, (b) Saturn, (c) Uranus, and (d) Neptune.  The qualitative
  features of the jets, which are the principle structures of the
  large-scale circulation on the Solar System giant planets, are
  captured for all four planets.  Only the observed parameters of
  Table~\ref{tab:one} are used, and no explicit thermal forcing is
  imposed.  The width of the jets roughly correspond to $L_\beta$.  On
  all four planets, \LR/$R_p\lsim 1/3$, which allows the formed jets
  to be stable over long time.  The sign of the equatorial jets on
  Jupiter and Saturn are reversed, compared with observations; but,
  the sign for Uranus and Neptune is as observed.  The scale for
  Jupiter is different from that for the others.}
\label{ss_jets}
\end{figure}  

\clearpage

\begin{figure}
\figurenum{7}
\epsscale{1.}
\caption{A ``warm'' Jupiter, with an equilibrium temperature, $T_e =
  300$~K.  Contour maps of the potential vorticity ($q$) field at
  $\tau\! =\! 199$ are shown in equatorial orthographic (a) and polar
  stereographic (b) views.  The physical parameters used are identical
  to the simulation in Figures~\ref{jupiter_side} and
  \ref{jupiter_top}, except that $\bar{\cal H}$ is 2.3 times larger
  (see text for explanation).  This case can serve as an example of a
  ``Jupiter'' which has migrated in to $\sim$1~AU from the central
  star.  Additionally, it could apply to a ``cool'', unsynchronized
  extrasolar giant planet with Jupiter $\Omega$ or to a level 2.3
  pressure scale heights deeper than in previous models (compare with
  Figures~\ref{jupiter_side} and \ref{jupiter_top}).  Due to the
  increased \LR\ value, the bands and jets are less pronounced, and
  the equatorial jet amplitude is reduced.  There are also fewer
  vortices in the polar region.}
\label{warmer_jupiter}
\end{figure}  

\clearpage

\begin{figure}
\figurenum{8}
\caption{Simulation of a \HDblah-like, unsynchronized close-in
  extrasolar giant planet in polar stereographic view: early
  evolution.  The rotation rate for this planet is that of \HDblah\
  (i.e., 8.4 times slower than that of Jupiter).  The global average
  root mean square velocity $\bar{U}$ for \HDblah\ is not known and
  can range plausibly between $\sim$10$^2$~m~s$^{-1}$ to
  $\sim$10$^3$~m~s$^{-1}$.  In this simulation, $\bar{U} =
  400$~m~s$^{-1}$, roughly the maximum value of all the planets in the
  Solar System.  We have varied this parameter in our study and
  summarize the findings in \S5.  The salient feature here is that the
  vortices and jets in the flow are much more dynamic compared with
  those in Figure~\ref{jupiter_top}, and even with those in
  Figure~\ref{neptune}.  This is due to the larger \LR\ value, which
  is $\sim\! R_p$ in this case.}
\label{unsync_hdblah_early}
\end{figure}  

\clearpage

\begin{figure}
\figurenum{9}
\caption{Same simulation and view of Figure~\ref{unsync_hdblah_early}:
  evolution {\it post} equilibrium.  By $\tau\! =\! 70$ a broad band
  of smoothly varying $q$ forms outside a robust polar vortex, which
  has formed through continuous mergers.  The flow is more similar to
  that on Neptune than on Jupiter (compare with
  Figures~\ref{jupiter_top} and \ref{neptune}).  As in Neptune, a
  polar vortex forms in each hemisphere, but the vortex is not
  centered on the pole.  Robust, off--pole vortices are general
  features in simulations of the circulation of \HDblah.  Another
  difference, consistent with less homogenized equatorial band, is the
  planetary (Rossby) wave breaking that appears more pronounced at
  mid-latitudes.}
\label{unsync_hdblah_late}
\end{figure}

\clearpage

\begin{figure}
\epsscale{.5}
\figurenum{10}
\caption{Quasi-steady jet profiles corresponding to the simulation of
  Figures~\ref{unsync_hdblah_early} and \ref{unsync_hdblah_late} (a)
  and the same simulation with a different $\bar{U}$, which is
  100~m~s$^{-1}$ (b).  The low number of jets is also a robust feature
  of our synchronized \HDblah\ simulations.  The precise profile of
  the jets depends on the value of $\bar{U}$, and the profile is
  slightly time-varying in a given run.  With a value of $\bar{U}$
  characteristic of most Solar System giant planets, three or four
  broad jets form.  When $\bar{U}\lsim 100$~m~s$^{-1}$ more jets may
  form.  Occasionally a two-jet profile forms for $\bar{U} \gsim
  400$~m~s$^{-1}$ and also for $\bar{U}\! =\! 0$ (see
  Figure~\ref{rest_forced}).  Note that in (b) the equatorial jet is
  prograde, demonstrating that a retrograde equatorial jet is {\it
    not} a necessary outcome of single-layer simulations.  That
  outcome requires a small \LR\ value.}
\label{unsync_hdblah_jets}
\epsscale{1.}
\end{figure}          

\clearpage

\begin{figure}
\figurenum{11}
\caption{Unsynchronized \HDblah\ simulations with different
  equilibrium temperatures, $\bar{T}_0$: (a)~1800~K and (b)~800~K.
  The two temperatures are chosen to roughly bracket a variety of
  physical situations (e.g., distance, stellar flux, and albedo).  For
  fixed flux and albedo, the higher $\bar{T}_0$ value can also
  represent an \HDblah-like planet which is $\sim$.005~AU closer to
  the host star than \HDblah\ is to its star.  Similarly, the lower
  $\bar{T}_0$ value can represent an \HDblah-like planet that is
  $\sim$0.15~AU further away.  The higher $\bar{T}_0$ value leads to a
  more smoothly varying low- to mid-latitude potential vorticity
  distribution, while the lower $\bar{T}_0$ value leads to much more
  planetary (Rossby) wave activity, along with a more centered polar
  vortex.  The basic three-jet profile is not changed in either case.}
\label{unsync_hdblah_temp}
\end{figure}  

\clearpage

\begin{figure}
\figurenum{12}
\caption{\HDblah\ atmospheric flow for different forcing amplitudes,
  $\eta$: (a) 0., (b) .02, (c) .04, and (d) .20.  All other parameters
  in (a)--(d) are identical; in particular, $\bar{U}\! =\!
  100$~m~s$^{-1}$.  Formally, $\eta$ can range from 0 to 1---i.e.,
  from no asymmetric forcing to full forcing, respectively.  The
  latter corresponds to a temperature deviation, $\delta T = {\cal
    O}(\bar{T}_0)$, which is not physically realistic on large scales.
  The primary effect of increasing the forcing asymmetry is to sharpen
  the midlatitude jet (a--c).  When the forcing is large, it induces
  large zonal asymmetry in the jet, destabilizing it and causing waves
  to break at high latitudes (d).  A simulation with $\eta\! =\! 0.4$
  shows a flow that is basically a more violent version of (d).}
\label{forcing_compare_flow}
\end{figure}  

\clearpage

\thispagestyle{empty}
\setlength{\voffset}{-30mm}

\begin{figure}
\figurenum{13}
\epsscale{.9}
\caption{Temperature distribution, $T_0 = T_0(\lambda,\phi)$, of the
  runs in Figure~\ref{forcing_compare_flow} with $\eta$: (a) 0., (b)
  .02, (c)~.04, and (d) .20.  For a fixed $\bar{U}$, $T_0$ is strongly
  dependent on $\eta$.  With no or low-amplitude forcing, $T_0$
  essentially tracks the flow structure (a).  As the forcing amplitude
  increases, however, there is a competition between the flow-induced
  and the forcing-induced temperature patterns: temperature anomalies
  either rotate quasi-steadily about the pole (b) or oscillate
  aperiodically about a point near the substellar point (c).  For high
  $\eta$ values, the forcing dominates, and $T_0$ is essentially the
  day-night distribution centered on the substellar point~(d).  In
  each case, a cyclone/anticyclone pair at high latitudes is present
  (i.e., independent of the amplitude) and is a robust feature of our
  simulations (see Figures~\ref{forcing_compare_flow} and
  \ref{cyclone_chop}). }
\label{forcing_compare_temp}
\epsscale{1.}
\end{figure}  

\clearpage

\setlength{\voffset}{0mm}

\begin{figure}
\figurenum{14}
\caption{\HDblah\ circulation for different global root mean square
  velocities, $\bar{U}$: (a) 100~m~s$^{-1}$, (b)~200~m~s$^{-1}$,
  (c)~400~m~s$^{-1}$, and (d) 1000 m s$^{-1}$.  The primary effect of
  larger $\bar{U}$ values is to strengthen and enlarge the polar
  vortex at high latitudes.  At low to mid latitudes, breaking of
  planetary (Rossby) waves appear to be reduced when $\bar{U}$ is
  larger.  However, in reality, the breaking also increases, as shown
  in Figure~\ref{cyclone_chop}.  The increase in vortex size is due to
  the initially more energetic vortices, which coalesce to form the
  final vortex.  The vortex in (d) is large enough to nearly cap the
  planet, poleward of 60 degrees latitude, if it were centered at the
  pole.}
\label{ubar_compare_flow}
\end{figure}  

\clearpage

\begin{figure}
\figurenum{15}
\caption{Flow outside the polar vortex and global jet profile (for two
  different $\bar{U}$ values).  Three same time frames of the
  simulation in Figure~\ref{ubar_compare_flow}d are shown in (a--c).
  There are 50 contours in each map.  Note the dominance of the
  cyclonic polar vortex (a) and the presence of a weaker, anticyclonic
  vortex, which is revealed when a smaller contour range is used (b).
  The anticyclone forms from high-amplitude breaking flow outside the
  cyclonic vortex (c).  The cyclone is associated with a coherent cold
  region, and the anticyclone is associated with a coherent warm
  region---as shown in Figure~\ref{ubar_compare_temp}.  Zonal jet
  profiles in two runs with different $\bar{U}$ values:
  (d)~100~m~s$^{-1}$ and (e)~1000~m~s$^{-1}$.  The number and width of
  the jets is qualitatively consistent with the Rhines scale,
  $L_\beta$. }
\label{cyclone_chop}
\end{figure}  

\clearpage

\thispagestyle{empty}
\setlength{\voffset}{-30mm}

\begin{figure}
\figurenum{16}
\epsscale{.9}
\caption{\HDblah\ atmospheric temperature distribution $T_0$ for
  different global root mean square velocity $\bar{U}$: (a) 100 m
  s$^{-1}$, (b)~200 m s$^{-1}$, (c) 400 m s$^{-1}$, and (d) 1000 m
  s$^{-1}$.  The same simulations as in Figure~\ref{ubar_compare_flow}
  are shown.  With forcing amplitude $\eta$ fixed, $T_0$ is strongly
  dependent on $\bar{U}$.  At low $\bar{U}$ values, the forcing
  dominates and $T_0$ is essentially the day-night distribution
  centered on the substellar point (a).  As $\bar{U}$ increases,
  competition between the forcing-induced and the flow-induced
  temperature patterns develops: temperature anomalies oscillate
  aperiodically about a point near the substellar point (b) or rotate
  quasi-steadily about the pole (c).  At large $\bar{U}$ values, $T_0$
  essentially tracks the flow structure on a several-days
  timescale~(d).}
\label{ubar_compare_temp}
\epsscale{1.}
\end{figure}  

\clearpage

\setlength{\voffset}{0mm}

\begin{figure}
\figurenum{17}
\caption{Simulation starting at rest.  The model is augmented to
  include the baroclinic effect of large-amplitude,
  zonally-asymmetric, equatorial forcing.  Potential vorticity ($q$)
  contours in cylindrical projection, centered at the substellar
  point, are shown for three time frames (a--c).  Corresponding zonal
  jet profiles are shown to the right of each $q$-map.  The forcing
  excites planetary (Rossby) waves (a).  Angular momentum is
  transported as propagating planetary waves interact with the
  background flow (b).  Eventually, the flow becomes very complex, at
  the high resolution of this simulation (c).  However, the
  north-south symmetry is maintained throughout, indicating the
  dominance of the forcing.}
\label{rest_forced}
\end{figure}  

\clearpage

\begin{figure}
\figurenum{18}
\caption{Summary of results with quiescent initial condition.  The
  plotting is identical to Figure~\ref{rest_forced}.  Late-time
  results in three simulations with different forcing conditions are
  shown: (a) day-night thermal forcing only, (b) with angular momentum
  forcing added, and (c) with initial random stirring added
  (i.e. atmosphere no longer at rest initially).  With only day-night
  forcing applied, the flow is laminar and consistent with $q$ lines
  distorted by the thermal forcing.  Finally, with angular momentum
  forcing also added, the flow becomes more complex and the atmosphere
  superrotates.  With the addition of initial random stirring, the
  flow becomes ``eddy driven'', if the stirring is energetic.  The
  equatorial jet is weakly retrograde in this case.}
\label{rest_compare}
\end{figure}

\end{document}